\documentclass[
 	reprint,
	amsmath,
    amssymb,
	aps,
    floatfix,
]{revtex4-1}

\usepackage[english]{babel}
\usepackage[utf8x]{inputenc}
\usepackage[T1]{fontenc}
\usepackage{braket}
\usepackage{mathtools}
\usepackage{flushend}

\usepackage{amsmath}
\usepackage{graphicx}
\usepackage{dcolumn}
\usepackage{bm}
\usepackage{array}
\usepackage[shortlabels]{enumitem}
\newcolumntype{?}{!{\vrule width 1pt}}
\usepackage[colorinlistoftodos]{todonotes}
\usepackage[colorlinks=true, allcolors=blue]{hyperref}

\usepackage{xcite}
\usepackage{bibunits}
\defaultbibliographystyle{apsrev4-1}
\usepackage{etoolbox}

\newcommand\textlcsc[1]{\textsc{\MakeLowercase{#1}}}

\makeatletter
\let\cat@comma@active\@empty
\makeatother

\begin{document}
\title{Topological Singularity Induced Chiral Kohn Anomaly in a Weyl Semimetal}
\author{Thanh Nguyen$^{1\dagger}$, Fei Han$^{1\dagger}$, Nina Andrejevic$^{2\dagger}$, Ricardo Pablo-Pedro$^{1\dagger}$, Anuj Apte$^{3}$, \\Yoichiro Tsurimaki$^{4}$, Zhiwei Ding$^{2}$, Kunyan Zhang$^{5}$,  Ahmet Alatas$^{6}$, Ercan E. Alp$^{6}$, Songxue Chi$^{7}$, Jaime Fernandez-Baca$^{7}$, Masaaki Matsuda$^{7}$, David Alan Tennant$^{7}$, Yang Zhao$^{8,9}$, Zhijun Xu$^{8,9}$, Jeffrey W. Lynn$^{8}$, Shengxi Huang$^{5}$, and Mingda Li$^{1*}$\\\hspace{0.1cm}}
\affiliation{\\
 $^{1}$Department of Nuclear Science and Engineering, MIT, Cambridge, MA, 02139, USA\\
 $^{2}$Department of Materials Science and Engineering, MIT, Cambridge, MA, 02139, USA\\
 $^{3}$Department of Physics, MIT, Cambridge, MA, 02139, USA\\
 $^{4}$Department of Mechanical Engineering, MIT, Cambridge, MA, 02139, USA\\
 $^{5}$Department of Electrical Engineering, The Pennsylvania State University, University Park, PA, 16802, USA\\
 $^{6}$Advanced Photon Source, Argonne National Laboratory, Lemont, IL, 60439, USA\\
 $^{7}$Neutron Scattering Division, Oak Ridge National Laboratory, Oak Ridge, TN, 37831, USA\\
 $^{8}$NIST Center for Neutron Research, National Institute of Standards and Technology, Gaithersburg, MD 20899, USA\\
 $^{9}$Department of Materials Science and Engineering, University of Maryland, College Park, Maryland, 20742, USA\\
 \newline
 $^\dagger$These authors contributed equally to this work.\\
 $^*$Corresponding author: mingda@mit.edu. \\
}
\date{\today}

\begin{bibunit}

\begin{abstract}
    The electron-phonon interaction (EPI) is instrumental in a wide variety of phenomena in solid-state physics, such as electrical resistivity in metals \cite{pines_nozieres_1999}, carrier mobility, optical transition and polaron effects in semiconductors \cite{ridley_2013,elliott_1957}, lifetime of hot carriers \cite{allen_1987,tisdale_2010, kozawa_2014}, transition temperature in BCS superconductors \cite{schrieffer_1999}, and even spin relaxation in diamond nitrogen-vacancy centers for quantum information processing \cite{wrachtrup_2006,fu_2009,jarmola_2012}. However, due to the weak EPI strength, most phenomena have focused on electronic properties rather than on phonon properties. One prominent exception is the Kohn anomaly, where phonon softening can emerge when the phonon wavevector nests the Fermi surface of metals \cite{kohn_1959}. Here we report a new class of Kohn anomaly in a topological Weyl semimetal (WSM), predicted by field-theoretical calculations, and experimentally observed through inelastic x-ray and neutron scattering on WSM tantalum phosphide (TaP). Compared to the conventional Kohn anomaly, the Fermi surface in a WSM exhibits multiple topological singularities of Weyl nodes, leading to a distinct nesting condition with chiral selection, a power-law divergence, and non-negligible dynamical effects. Our work brings the concept of Kohn anomaly into WSMs and sheds light on elucidating the EPI mechanism in emergent topological materials.
\end{abstract}

\maketitle


\noindent In 1959, Walter Kohn proposed the anomalous phonon dispersion behavior in a metal, which arises when electrons lose their dielectric screening \cite{kohn_1959}. This anomaly, known as a Kohn anomaly, directly images the Fermi surface on the phonon dispersion, and overturned the long belief that the weak EPI can only lead to negligible effects on phonon properties. Intuitively, a Kohn anomaly occurs when electronic states $\boldsymbol{k}_1$ and $\boldsymbol{k}_2$ near the Fermi surface are parallelly nested by a phonon with wavevector $\boldsymbol{q}\equiv\boldsymbol{k}_2-\boldsymbol{k}_1$. This is a stringent condition only met by a single $q\equiv|\boldsymbol{q}|$ value at $q\approx 2k_F$, where $k_F$ is the Fermi wavevector. Extensive research has delved into the role of Kohn anomalies in conventional \cite{aynajian_2008} and unconventional superconductors \cite{weber_1987,reznik_2006}, carbon materials such as carbon nanotubes \cite{piscanec_2007}, graphene \cite{mafra_2009,tse_2008,wunsch_2006}, and graphite \cite{ferrari_2007}, as well as other low-dimensional systems such as 1D conductors \cite{renker_1973} and topological insulators \cite{zhu_2012}.

The recent development of WSMs \cite{xu_2015,lv_2015,weng_2015,huang_2015,min_2019} offers a new platform to realize  exotic phonon properties, such as the phonon Hall effect \cite{cortijo_2015}, chiral magnetic effect \cite{song_2016} and chiral anomaly in phonon spectra \cite{rinkel_2017}. As such, WSMs could serve as an intriguing platform to study Kohn anomalies due to the presence of topologically protected Weyl nodes and 3D linear-dispersive Weyl fermions.

In this \textit{Letter}, we theoretically predict and experimentally observe a new type of Kohn anomaly in WSM, which exhibits a few novel features. First, since the simply-connected Fermi surface in a conventional Fermi liquid evolves into disconnected topological singularities of chiral Weyl nodes, the condition to achieve Kohn anomalies becomes largely relaxed. As the EPI does not change chirality, it plays an essential role in understanding the coupling strength through the following requisite. For two Weyl nodes located at $\boldsymbol{k}_{W1}$ and $\boldsymbol{k}_{W2}$, both which necessarily share the same chirality, a phonon with $\boldsymbol{q}\approx\boldsymbol{k}_{W2}-\boldsymbol{k}_{W1}$ can directly lead to the anomaly. In particular, for our chosen material of type-I WSM TaP, which contains two sets of inequivalent Weyl nodes $\{\boldsymbol{k}_{W1}\}$ and $\{\boldsymbol{k}_{W2}\}$, there is a subset of nodes satisfying $\boldsymbol{k}_{W2}-\boldsymbol{k}_{W1}\in\{\boldsymbol{k}_{W2}\}$ \cite{weng_2015}. As a result, phonon $\boldsymbol{q}$ values can be chosen directly at Weyl nodes. Second, instead of a logarithmic divergence as in a Fermi liquid or a weak power-law divergence as in graphene \cite{hwang_2008}, the Kohn anomaly in a WSM shows a stronger power-law divergence. This counterintuitive result originates from the 3D dispersion. Third, in WSM, the Debye frequency $\omega_D$ can be higher than the Fermi level $E_F$, indicating a non-negligible dynamical effect, since the frequency dependence of the dielectric function occurs on the scale of $\omega_D$. Such a dynamical effect leads to softening within a finite regime in the Brillouin zone instead of at an individual $\boldsymbol{q}$ point. This contrasts with the conventional Fermi liquid, where static screening suffices since $\omega_D \ll E_F$. Our work represents the first reported observation of a Kohn anomaly in a topological nodal semimetal, and offers a new tool for probing the EPI characteristics in a broader category of topological materials.

\textit{Field-theoretical calculation of topological Kohn anomaly.} The dielectric screening from inter-node scattering via a phonon $\boldsymbol{q}=\boldsymbol{k}_{W2}-\boldsymbol{k}_{W1}$ is computed for 3D linear dispersive Weyl fermions using $E(\boldsymbol{k})=\pm v_F|\boldsymbol{k}-\boldsymbol{k}_{Wj}| - \mu_j$, where $v_F$ is the Fermi velocity and $j=1,2$ denotes different Weyl nodes at chemical potential $\mu_j$. The polarization operator $\Pi(\nu,\boldsymbol{q})$ from the inter-Weyl-node scattering can be written as (Supplementary A \cite{a_note1}):
\begin{equation}
\resizebox{0.9\linewidth}{!}{$
\Pi(\nu, \boldsymbol{q}) \approx -\frac{\left(v_F q^{\prime}\right)^{2}}{24 \pi^2}\left(\ln \left|\frac{4\left(v_F \Lambda\right)^{2}}{\left(v_F q^{\prime}\right)^{2}-\nu^{\prime 2}}\right|+\frac{(v_Fq^{\prime})^2 - \nu^{\prime 2}}{10(v_F\Lambda)^2}-\frac{1}{3}\right)$}
\label{eq:polarization}
\end{equation}

\noindent where $q^\prime=|\boldsymbol{q}+\boldsymbol{k}_{W1}-\boldsymbol{k}_{W2}|$, $\nu^\prime = \nu-\mu_1+\mu_2$ and $\Lambda$ is a momentum cutoff. Since the dynamical dielectric function $\epsilon_r(\nu,\boldsymbol{q})$ can be written as $\epsilon_r(\nu,\boldsymbol{q}) = 1 - 4\pi e^2\Pi(\nu,\boldsymbol{q})/(\kappa q^2)$, we see immediately that a Kohn anomaly with a strong power-law divergence can occur with the following condition
\begin{equation}
   v_Fq^\prime = \pm \nu^\prime.
\end{equation}

\noindent by satisfying which the momentum derivative of dielectric function diverges:
\begin{equation}
    \frac{\partial \varepsilon_{r}}{\partial q}\propto \frac{\left(v_F q^{\prime}\right)^{3}}{\left(v_F q^{\prime}\right)^{2}-(\nu^{\prime})^2}.
\end{equation}

\begin{figure}[ht]
	\centering
	\includegraphics[width=\linewidth]{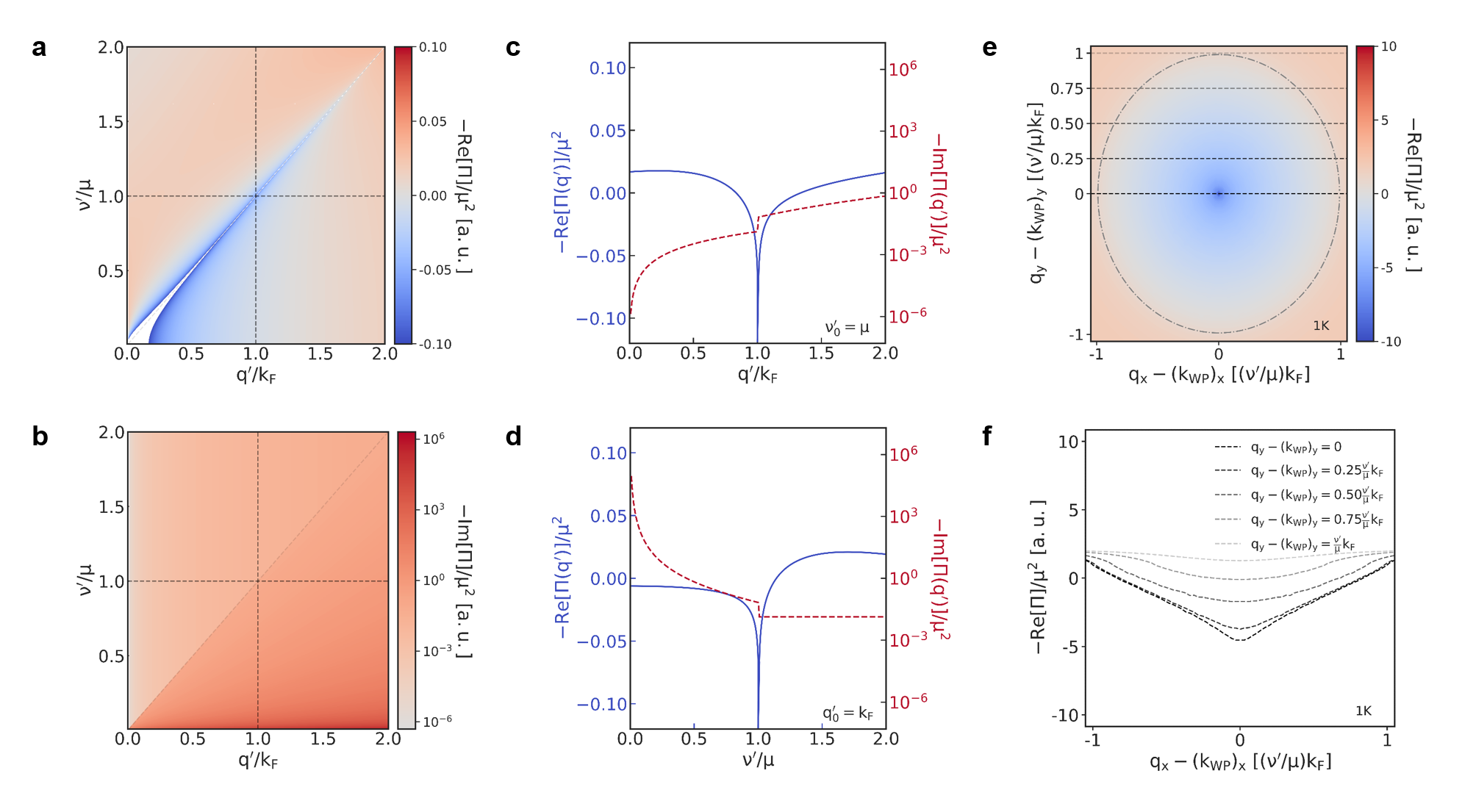}
	\caption{\textbf{Field-theoretical calculations of topological Kohn anomaly.} Density plot of the \textbf{a}, real, and the \textbf{b}, imaginary part of the polarization function $-\Pi(\boldsymbol{q}^{\prime}, \nu^{\prime})$ at $T=1K$ \textbf{c}. Constant-frequency line profile of  $-\textrm{Re}[\Pi(\boldsymbol{q}^{\prime}, \nu_0^\prime=\mu)]$ and $-\textrm{Im}[\Pi(\boldsymbol{q}^{\prime}, \nu_0^\prime=\mu)]$ \textbf{d}. Constant-wavevector line profile of  $-\textrm{Re}[\Pi(q_0^\prime = k_F, \nu^{\prime})]$ and $-\textrm{Im}[\Pi(q_0^\prime = k_F, \nu^{\prime})]$. \textbf{e}. Density plot of $-\textrm{Re}[\Pi(\boldsymbol{q_x}-\boldsymbol{k_{\text{WP}_x}}, \boldsymbol{q_y}-\boldsymbol{k_{\text{WP}_y}})]$, which represents a realistic experimental setup scanning near the Weyl node. \textbf{f} Line cuts at different values of deviations away from a Weyl node. There is a noticeable dip at Weyl node, yet even away from the Weyl nodes, softening can still exist due to the point-like Weyl node and dynamic effect.}
	\label{fig:1}
\end{figure}

The divergence condition for a Kohn anomaly implies that the momentum mismatch $\delta\boldsymbol{q}\equiv\boldsymbol{q}-(\boldsymbol{k}_{W2}-\boldsymbol{k}_{W1})$ can be compensated by a finite dynamical effect $\nu^\prime$. Consequentially, a patch of $\boldsymbol{q}$ values in momentum space with small momentum mismatch can still experience a Kohn anomaly. In fact, the simple divergence condition $v_Fq^\prime=\pm\nu^\prime$ will persist even with finite doping. The density plots of the real and imaginary parts of $-\Pi(\nu,\boldsymbol{q})$ at finite doping and temperature are shown in Figures \hyperref[fig:1]{1a} and \hyperref[fig:1]{1b} (additional temperatures in Figures S1 and S2). The divergence appears as a sharp peak along the line $v_Fq^\prime = \nu^\prime$, and is further visualized along constant-frequency $\nu_0^\prime/\mu = 1$ and constant-wavevector $q_0^\prime/k_F=1$ cuts in Figures \hyperref[fig:1]{1c} and \hyperref[fig:1]{1d}, respectively in reduced dimensionless units, where $\mu\equiv\sqrt{\mu_1^2+\mu_2^2}$, and $k_F\equiv\mu/v_F$. Figure \hyperref[fig:1]{1e} is a plot of $-\text{Re}[\Pi(\nu,\boldsymbol{q})]$ in $\boldsymbol{\delta q}_x$--$\boldsymbol{\delta q}_y$ space integrated from 0 to $\nu^{\prime}$, which is proportional to the magnitude of the phonon softening and  reveals a negative contribution emanating from the zero-mismatch condition as a result of the divergence at $v_Fq^\prime=\pm\nu^\prime$. The divergence is alleviated here by a small numerical imaginary part mimicking the considerations of additional scattering terms. Line cuts in $\boldsymbol{\delta q}_x$--$\boldsymbol{\delta q}_y$ space shown in Figure \hyperref[fig:1]{1f} reveal a broad dip in $-\text{Re}[\Pi(\nu, \boldsymbol{q})]$ that does not necessarily occur at the zero-mismatch condition, thereby demonstrating the existence of a non-negligible dynamical effect. It is worthwhile mentioning that although the polarization in Figure \hyperref[fig:1]{1c}  and \hyperref[fig:1]{1d} shows a sharp divergence, the distinct condition to fulfill a Kohn anomaly in a patch of $\boldsymbol{q}$ values eventually leads to a broad "bowl-shaped" softening in the Brillouin zone, which qualitatively agrees well with the experiments without knowing the EPI coupling constants. The EPI, embodied through an expression for $\Pi(\nu, \boldsymbol{q})$ characteristic of WSMs, results in an distinct renormalization of the bare ionic phonon frequencies within a range of $\boldsymbol{q}$-space conducive to a different nature of Kohn anomaly.

\begin{figure}[ht]
	\centering
	\includegraphics[width=\linewidth]{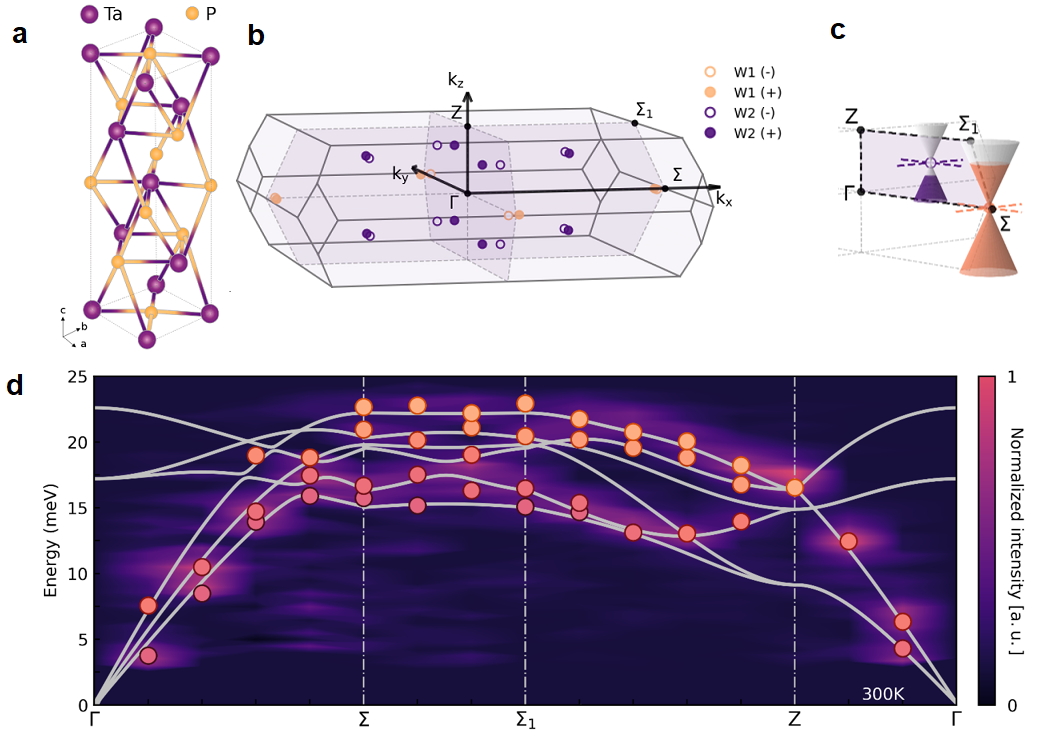}
	\caption{\textbf{Phonon dispersion of TaP.} \textbf{a}. Crystal structure of TaP with \textbf{b}, corresponding Brillouin zone and featuring the locations and Berry curvature signs of paired Weyl nodes W1 (orange) and W2 (purple). \textbf{c}. Schematic of relative energy location of W1 and W2 Weyl nodes. \textbf{d}. Low-energy phonon dispersion of TaP, measured by IXS and INS at 300K along the high-symmetry loop $\Gamma$-$\Sigma$-$\Sigma_1$-Z-$\Gamma$. Points represent extracted phonon modes from measurements. The grey lines denote \textit{ab initio} calculations showing excellent agreement, and the color map represents the spectra intensity.}
	\label{fig:2}
\end{figure}

\textit{Phonon dispersion along high-symmetry directions.} TaP crystallizes in the body-centered tetragonal space group I4$_1$md (109) (Figure \hyperref[fig:2]{2a}) and hosts two sets of inequivalent Weyl nodes, denoted W1 and W2 (Figure \hyperref[fig:2]{2b}). We first present the phonon dispersion measurements of a TaP crystal (Figure S3, Supplementary B and C \cite{b_note2}) along a high-symmetry loop $\Gamma$-$\Sigma$-$\Sigma_1$-Z-$\Gamma$ (Figure \hyperref[fig:2]{2c} with data in Figure S4) using inelastic X-ray scattering (IXS). We focused on the low-energy phonons ($<$25 meV), which include the Ta optical phonons but not those associated with the motion of P atoms. The phonon energies are extracted using damped harmonic oscillator models to convolute with the instrument resolution functions, and then fitting the measured spectra. The resulting phonon dispersion is shown in Figure \hyperref[fig:2]{2d} (and Figure S6 for lower temperatures), along which the intensity of fitting the intrinsic scattering (after deconvolution) is plotted as a color map. Grey lines designate \textit{ab initio} phonon dispersion calculations for which the procedure is detailed in Supplementary D \cite{c_note3}. Further data were collected using inelastic neutron scattering (INS), which agrees well with the dispersion from IXS (Figure S7). The excellent agreement between experiments and \textit{ab initio} calculations indicates a level of reliability of computational phonon spectra, which serves as a basis to compare phonon dispersions away from high-symmetry lines.

\begin{figure}[ht]
	\centering
	\includegraphics[width=\linewidth]{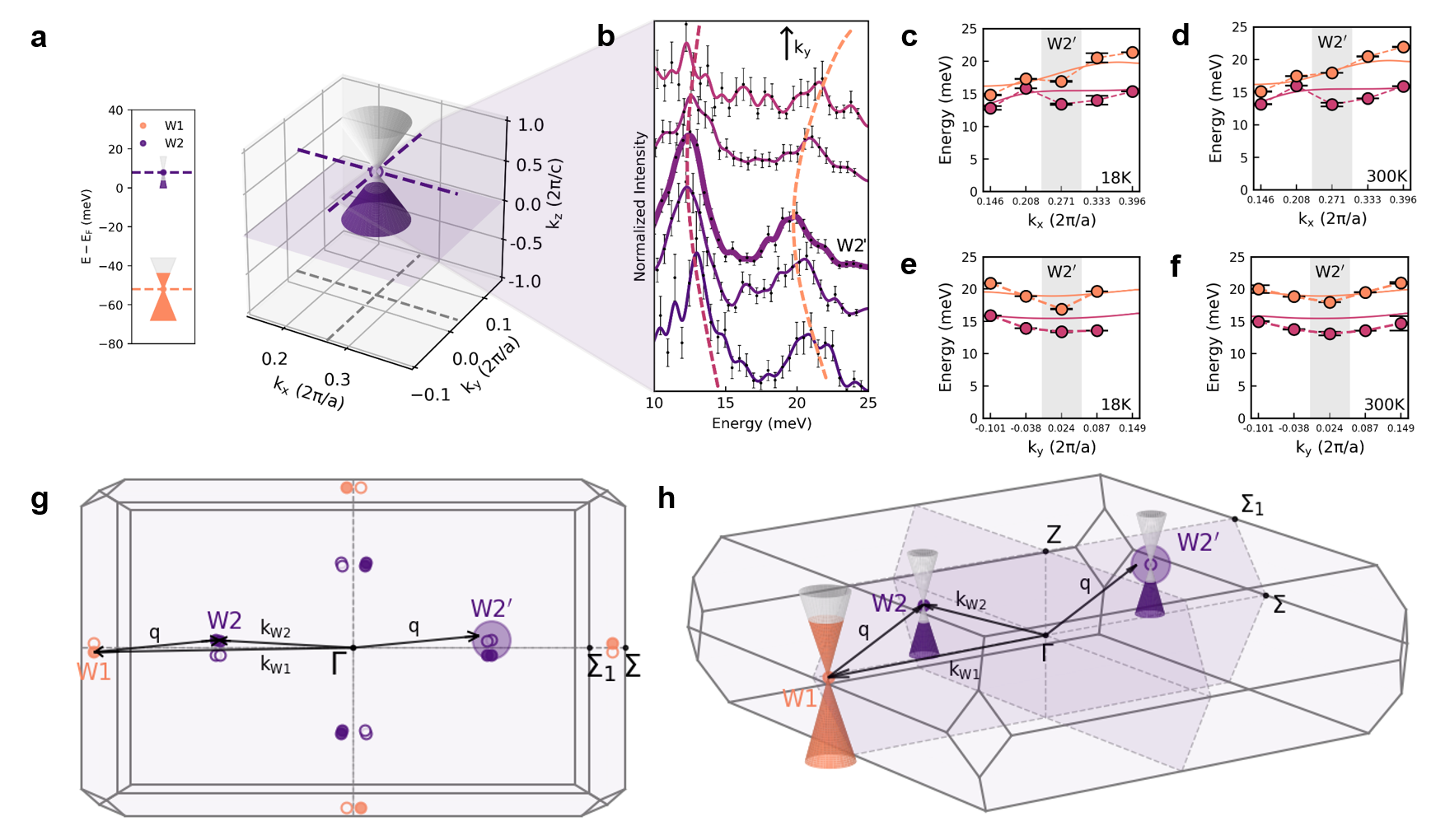}
	\caption{\textbf{Topological Kohn anomaly at W2 Weyl node.} \textbf{a}. Location of the W2 Weyl node in energy and momentum space. \textbf{b}. Phonon spectra near the W2 Weyl node, with central thicker line denoting the location of W2$^{\prime}$. Strong phonon softening at this Weyl node is observed. \textbf{c-f.} Dispersion of two representative phonon modes near W2 at 18K and 300K, along $k_x$ and $k_y$ directions. Solid lines correspond to \textit{ab initio} calculations without the EPI, where their discrepancy with experimental data further supports the strong softening near the W2 node.  2D (\textbf{g}) and 3D (\textbf{h}) showing the realization condition of Kohn anomaly, where a phonon  $\boldsymbol{q}=\boldsymbol{k}_{W2}-\boldsymbol{k}_{W1} \approx \boldsymbol{k}_{W2^{\prime}}$ connects two Weyl nodes $\boldsymbol{k}_{W1}$ and $\boldsymbol{k}_{W2}$. }
	\label{fig:3}
\end{figure}

\textit{Observation of topological chiral Kohn anomaly.} We present experimental signatures highlighting the presence of a Kohn anomaly at the W2 Weyl node, predicted in Figure \ref{fig:1}. In our TaP sample, the W1 Weyl nodes are $\sim60$meV below $E_F$ , while W2 nodes are a few meV above $E_F$ (Figure \hyperref[fig:3]{3a}). As a result, W1 represents a much larger carrier pocket than W2. We carried out IXS measurements at $\boldsymbol{q}$ values near a W2 Weyl node located at $\boldsymbol{k}_{W2^{\prime}}=(0.271,0.024,0.578)$ (Figure S8). Even without fitting, the phonon softening at the Weyl node can be seen clearly from the original data in Figure \hyperref[fig:3]{3b}. The bowl-shaped softening characteristics resembles the field-theoretical prediction in Figure \hyperref[fig:1]{1f} very well, although a quantitative agreement is unpractical, requiring the mode-resolved EPI contants. The fitted phonon dispersion relation of the highest acoustic and the lowest optical phonons along the $k_x$ and $k_y$ directions within the plane containing $\boldsymbol{k}_{W2^{\prime}}$ are shown in Figures \hyperref[fig:3]{3c}-\hyperref[fig:3]{3f}. Strong phonon softening is observed at both $T=18$K and $T=300$K, and in both $k_x$ and $k_y$ directions. Such softening is absent in \textit{ab initio} calculations without considering EPI. The semi-quantitative agreement between theory and experimental trends, the consistent softening behavior at multiple $\boldsymbol{q}$ points, and the absence of softening in \textit{ab initio} calculations without EPI consideration, overall strongly suggest an EPI nature of the phonon softening. Additionally, the phonon softening takes place at all measured temperatures, indicating a possible topological robustness. Such softening can be understood as a Kohn anomaly from inter-Weyl node scattering. In fact, it is possible to nest a W1 electronic state at $\boldsymbol{k}_{W1} = (-0.518,-0.014,0)$ with another W2 state $\boldsymbol{k}_{W2}=(-0.271,0.024,0.578)$, both with "+" chirality, via a phonon $\boldsymbol{q}=\boldsymbol{k}_{W2}-\boldsymbol{k}_{W1}\approx\boldsymbol{k}_{W2^{\prime}}$, with a mismatch $|\boldsymbol{q}-\boldsymbol{k}_{W2^{\prime}}|/|\boldsymbol{k}_{W2'}|\sim$ 4\%. Details relating to different nesting combinations are listed in Supplementary E. The schematics of this nesting condition are shown in Figures \hyperref[fig:3]{3g}-\hyperref[fig:3]{3h}. As mentioned previously, the dynamical effect significantly reduces the mismatch down to $\sim$ 1\% and further enables the Kohn anomaly to manifest at $\boldsymbol{q}\simeq\boldsymbol{k}_{W2'}$. In particular, the dynamical correlation almost exactly reproduces the strong phonon softening feature at Weyl node $\boldsymbol{k}_{W2^{\prime}}$: with $v_F\sim 1.5\times 10^5$m/s in TaP and the momentum mismatch being $v_F\delta\boldsymbol{q}\sim 50$meV, $\nu^\prime\sim 60$meV largely compensates for the mismatch, thereby facilitating the satisfaction of the divergence condition and agreeing with the observation in Figures \hyperref[fig:3]{3c}-\hyperref[fig:3]{3f}.

\begin{figure}[ht]
	\centering
	\includegraphics[width=\linewidth]{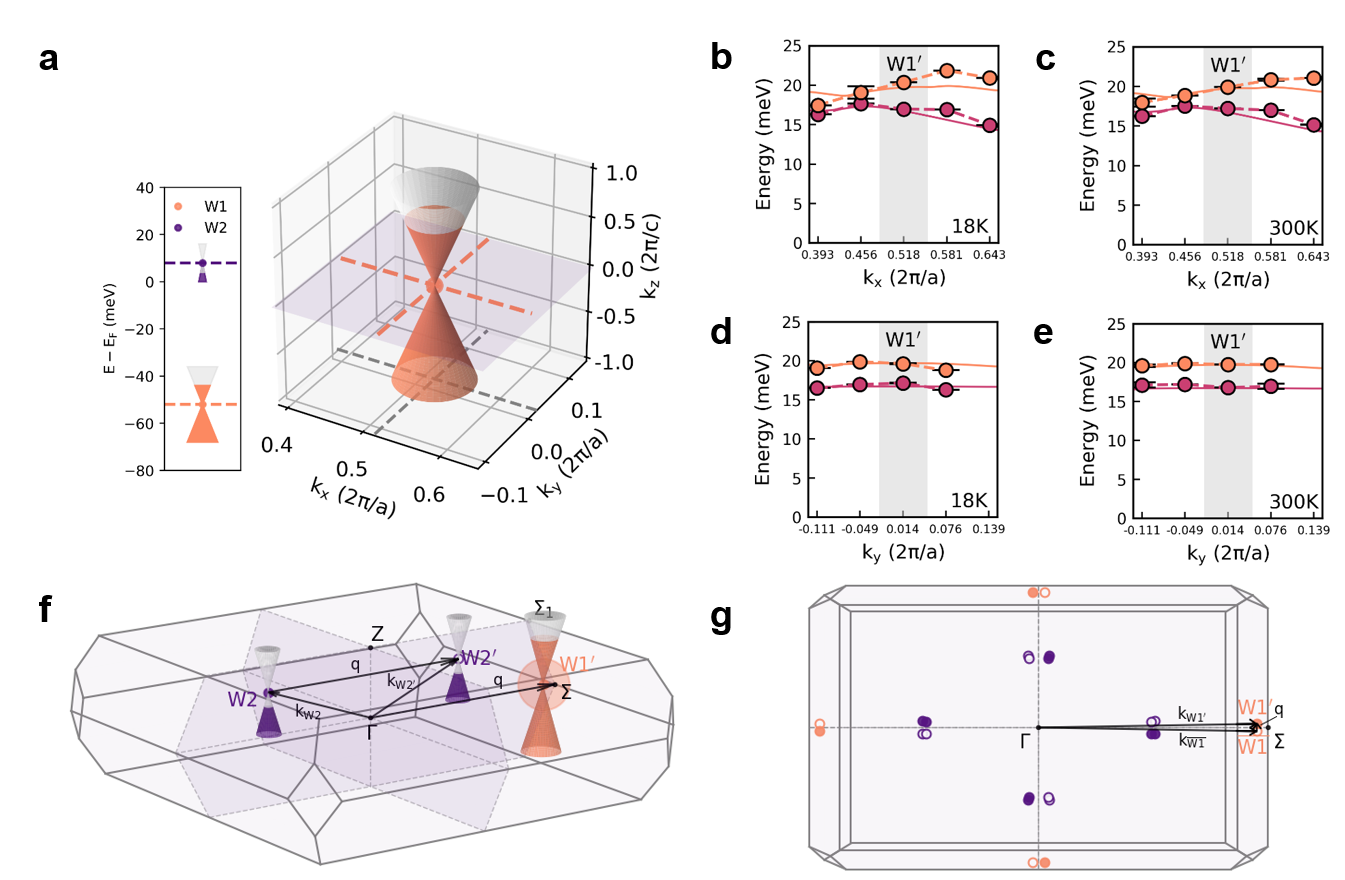}
	\caption{\textbf{Phonon characteristics at the W1 Weyl node and $\Gamma$-point.} \textbf{a}. Location of the W1 Weyl node in energy and momentum space. Dispersions of two representative phonon modes obtained from IXS data near W1 at \textbf{b}, 18K and \textbf{c}, 300K along the $k_x$ direction as well as \textbf{d} and \textbf{e}, along the $k_y$ direction. Error bars represent one standard deviation. Experimental phonon dispersion agrees excellently with \textit{ab initio} calculations (solid lines). \textbf{f}. Schematic of the connection of two equivalent electronic states at $\boldsymbol{k}_{W2}$ and $\boldsymbol{k}_{W2^{\prime}}$ by the phonon $\boldsymbol{q}=\boldsymbol{k}_{W2^{\prime}}-\boldsymbol{k}_{W2}\approx\boldsymbol{k}_{W1^{\prime}}$, the W1 node near which IXS measurements were performed. \textbf{g}. Schematic seen from the (001) plane of the connection between two equivalent electronic states at $\boldsymbol{k}_{W1^{\prime}}$ and $\boldsymbol{k}_{\overline{W1}}$ by the phonon momentum $\boldsymbol{q}=\boldsymbol{k}_{W1^{\prime}}-\boldsymbol{k}_{\overline{W1}}\approx \Gamma$.}
	\label{fig:4}
\end{figure}

\textit{Chirality selection of the topological chiral Kohn anomaly.} The IXS measurements carried out near a W1 Weyl node (Figures \hyperref[fig:4]{4a} and S9) present a contrasting result. When the phonon $\boldsymbol{q}$ is near a W1 node $\boldsymbol{k}_{W1^{\prime}}=(0.518,0.014,0)$, there is no clear indication of phonon softening at all measured temperatures, where measured phonon dispersions agree very well with \textit{ab initio} calculations in Figures \hyperref[fig:4]{4b}-\hyperref[fig:4]{4e}). This is largely due to a lack of a scattering channel that can simultaneously conserve momentum and chirality. For momentum-conserved scattering, although the phonon nesting condition can still roughly be met by considering scattering from $\boldsymbol{k}_{W2}$ to $\boldsymbol{k}_{W2^{\prime}}$, where $\boldsymbol{q}=\boldsymbol{k}_{W2^{\prime}}-\boldsymbol{k}_{W2}=(0.542,0,0)\approx\boldsymbol{k}_{W1^{\prime}}$ (mismatch $|\boldsymbol{q}-\boldsymbol{k}_{W1^{\prime}}|/|\boldsymbol{k}_{W1^{\prime}}| \sim$ 5\%) (Figure \hyperref[fig:4]{4f}), the W2 and W2$^{\prime}$ nodes have opposite chirality ("$+$" and "$-$", respectively), prohibiting the EPI to occur. On the other hand, for a chirality-conserved scattering $\boldsymbol{q}=\boldsymbol{k}_{W2^{\prime}}-\boldsymbol{k}_{\overline{W2} }$, where $\boldsymbol{k}_{\overline{W2}}=(-0.271,-0.024,0.578)$ gives the $\overline{\textrm{W2}}$ node paired with W2$^{\prime}$ and has "$-$" chirality, the $\sim 8\%$ momentum mismatch is simply too large to compensate even with dynamical effects considered. Moreover, the low carrier concentration of the hole pockets at W2 nodes further decreases the overall EPI contribution. As such, the magnitude of the Kohn anomaly at W1 should be imperceptible relative to the results at W2. This analysis corroborates the IXS data.

In addition to the two types of Weyl nodes, one may expect a Kohn anomaly to emerge at the $\Gamma$ point. However, the chirality-conserved scattering channel (say, W1 and W1$^{\prime}$) has large momentum mismatch away from the $\Gamma$-point, while the momentum-conserved channel (say, W1$^{\prime}$ and $\overline{\textrm{W1}}$ with $\boldsymbol{k}_{\overline{W1}} = (0.518,-0.014,0)$ does not preserve chirality as shown in Figures S11. A mixed behavior is observed from Raman scattering measurements. A weak softening $\sim$ 1.2meV is observed for the lower optical phonon at the $\Gamma$-point, but not for the higher-energy optical phonon. As a result, the intra-node scattering may still happen and requires further investigation.

To summarize, we theoretically formulated and experimentally demonstrated a Kohn anomaly in a WSM, which can lead to the anomalous broad-range phonon softening behaviors arising from the scattering between the topological singularity of chiral Weyl nodes. Unlike a conventional Fermi liquid, in TaP, with 8 W1 nodes and 16 W2 nodes, numerous $\boldsymbol{q}$ regimes in the Brillouin zone can achieve Kohn anomalies. The recent \textit{ab initio} calculations in Dirac semimetals also confirm the existence of Kohn anomalies in topological semimetals \cite{yue_2019,yue_2020}. Moreover, in contrast to the conventional Fermi liquid with only static screening and logarithmic divergence $d\epsilon_r/dq \propto \ln(|q-2k_F|/4k_F)$, or graphene and 2D electron gas with $\partial\epsilon_r/\partial q\propto1/\sqrt{q^2-4k_F^2}$, we find a highly distinct divergence condition at $v_Fq^\prime \rightarrow \pm \nu^\prime$ with a leading term $\partial\epsilon_r/\partial q \propto (v_Fq^{\prime})^3/[(v_Fq^\prime)^2 - \nu^{\prime 2}]$, where dynamical effects play a critical role with explicit Weyl node location dependence and chirality selection. The Kohn anomaly identified in our work highlights a previously overlooked instance of EPI in WSMs, and can offer a ubiquitous tool to extract the EPI strength, as carried out in graphite \cite{piscanec_2004}, to elucidate the interplay between chiral Weyl fermions and phonons. Our discovery adds to the rich array of exotic EPI effects realized in novel topological materials.

\section*{Acknowledgments}
The authors thank S.Y. Xu for the helpful discussions. T.N. thanks the support from the MIT SMA-2 Fellowship Program and MIT Sow-Hsin Chen Fellowship. N.A. acknowledges the support of the National Science Foundation Graduate Research Fellowship Program under Grant No. 1122374.  R.P.P. thanks the support from FEMSA and ITESM. A. Apte thanks the support of MIT John Reed fund. Y.T. and Z.D. thank the support from DOD Defense Advanced Research Projects Agency (DARPA) Materials for Transduction (MATRIX) program under Grant HR0011-16-2-0041. D.A.T. was sponsored by the Laboratory Directed Research and Development Program (LDRD) of Oak Ridge National Laboratory, managed by UT-Battelle, LLC, for the U.S. Department of Energy (Project ID 9533). This research used resources of the Advanced Photon Source, a U.S. Department of Energy (DOE) Office of Science User Facility operated for the DOE Office of Science by Argonne National Laboratory under Contract No. DE-AC02-06CH11357. This research on neutron scattering used neutron research facilities at the High Flux Isotope Reactor, a DOE Office of Science User Facility operated by the Oak Ridge National Laboratory and at the NIST Center for Neutron Research (NCNR), at the National Institute of Standards and Technology, an agency of the U.S. Department of Commerce. The identification of any commercial product or trade name does not imply endorsement or recommendation by the National Institute of Standards and Technology. M.L. acknowledges the neutron sample alignment support from MIT Nuclear Reactor Laboratory Seed Fund Program. T.N., N.A., F.H. and M.L. acknowledge the support from U.S. Department of Energy (DOE), Office of Science (SC), Basic Energy Sciences (BES), award No. DE-SC0020148.

\putbib[references]

\end{bibunit}

\onecolumngrid

\clearpage
\onecolumngrid

\clearpage

\clearpage

\clearpage

\section*{Supplementary materials for Topological Singularity Induced Chiral Kohn Anomaly in A Weyl Semimetal}
\setcounter{page}{1}
\begin{bibunit}

\setcounter{equation}{0}
\setcounter{table}{0}
\renewcommand{\theequation}{S\arabic{equation}}
\renewcommand{\thetable}{S\arabic{table}}

\nocite{pines_nozieres_1999}
\nocite{ridley_2013}
\nocite{elliott_1957}
\nocite{allen_1987}
\nocite{tisdale_2010}
\nocite{kozawa_2014}
\nocite{schrieffer_1999}
\nocite{wrachtrup_2006}
\nocite{fu_2009}
\nocite{jarmola_2012}
\nocite{kohn_1959}
\nocite{aynajian_2008}
\nocite{weber_1987}
\nocite{reznik_2006}
\nocite{piscanec_2007}
\nocite{mafra_2009}
\nocite{tse_2008}
\nocite{wunsch_2006}
\nocite{ferrari_2007}
\nocite{renker_1973}
\nocite{zhu_2012}
\nocite{xu_2015}
\nocite{lv_2015}
\nocite{weng_2015}
\nocite{huang_2015}
\nocite{min_2019}
\nocite{cortijo_2015}
\nocite{song_2016}
\nocite{rinkel_2017}
\nocite{hwang_2008}
\nocite{a_note1}
\nocite{b_note2}
\nocite{c_note3}
\nocite{yue_2019}
\nocite{yue_2020}
\nocite{piscanec_2004}
\nocite{kapusta_2006}
\nocite{bjorken_1966}
\nocite{lv_2013}
\nocite{zhou_2015}
\nocite{thakur_2018}
\nocite{sadhukhan_2020}
\nocite{han_2019}
\nocite{sinn_2001}
\nocite{alatas_2011}
\nocite{toellner_2011}
\nocite{lynn_2012}
\nocite{dave_2009}
\nocite{kresse_1999}
\nocite{kresse_1996}
\nocite{kresse_1995}
\nocite{perdew_1996}
\nocite{togo_2015}

\section*{Contents}
\begin{enumerate}[(A)]
    \item Derivation of polarization and dielectric functions for Weyl semimetals
    \item Single crystal growth and sample preparation of TaP for IXS
    \item X-ray and neutron scattering measurements and analysis of TaP
    \item Computational details
    \item Phonon nesting conditions between Weyl nodes in TaP
\end{enumerate}

\subsection*{Supplementary A: Derivation of polarization and dielectric functions for Weyl semimetals}
\label{sec:supplementaryA}
\subsection{Introduction}   
\noindent In the study of graphene, the dynamical polarization function $\Pi(\nu,\mathbf{q})$ is one of the fundamental quantities required to understand physical properties of the system such as screening due to a charged impurity and the existence of collective excitations such as plasmons. Prior to its investigation in a Weyl semimetal, $\Pi(\nu,\mathbf{q})$ had been extensively studied for graphene. In this section of the paper, we examine in greater detail the dynamical polarization function for a Weyl semimetal which has the possibility of exhibiting topological Fermi arcs on the surface and as well as phenomena such as the chiral magnetic effect in the bulk.\\ 

\noindent To derive the polarization function $\Pi(\nu,\mathbf{q})$, we will consider a simple continuous model which can be used to describe Weyl semimetals and nodal-line semimetals. Expanding around the $\Gamma$ point, we consider a Hamiltonian in momentum space with tilted Dirac cones as the following
\begin{equation}
    \mathcal{H}(\mathbf{k}, \mathbf{d})=\hbar v_{F}\left(\mathbf{k} \cdot \sigma+ p\mathbf{d} \cdot \mathbf{k} I_{2}\right)  -\mu I_2 \label{Hamiltonian}
\end{equation}

\noindent where $\mathbf{k}=(k_x,k_y,k_z)$ is the momentum and $\sigma=(\sigma_x,\sigma_y,\sigma_z)$ are  the Pauli matrices. In addition, $v_F$ represents the Fermi velocity, $\mathbf{d}$ is a unit vector,  $p$ is a dimensionless parameter which will be referred to as the \textit{tilting parameter}, and $\mu$ represents the chemical potential. The Hamiltonian above has the following eigenvalues
\begin{equation}
    E_{\pm}=\hbar v_F(p\mathbf{d} \cdot \mathbf{k} \pm  |\mathbf{k}|)-\mu
\end{equation}

\noindent Near the band crossing at the Weyl point, the low-energy bands can be linearized in the variation $\mathbf{k}'=\mathbf{k}-\mathbf{k}_{Wi}$ of the $\mathbf{k}$ vector with respect to the position $\mathbf{k}_{Wi}$ of the Weyl node $Wi$ which in our case, $i \in \{1,2\}$. Note that the subscript $i$ does not necessarily refer to the subset of inequivalent Weyl nodes as it does in the main text. As such, we can write the Hamiltonian as follows 
\begin{equation}
    \label{eq:Hamiltonian}
    \mathcal{H}(\mathbf{k}^{\prime}, \mathbf{d})=\hbar v_{F}\left(\mathbf{k}^{\prime} \cdot \sigma+ p\mathbf{d} \cdot \mathbf{k}^{\prime} I_{2}\right)  -\mu I_2
\end{equation}

\noindent with eigenvalues 
\begin{equation}
    E_{\pm}=\hbar v_F(p\mathbf{d} \cdot \mathbf{k}^{\prime} \pm  |\mathbf{k}^{\prime}|)-\mu
\end{equation}

\noindent The propagator is given by 
\begin{equation}
    G(i\omega_n, \mathbf{k}^{\prime}, \mathbf{d})= \frac{1}{(i \omega_n +\mu - \hbar v_F p \mathbf{d}\cdot \mathbf{k}^{\prime})I_2 -  \hbar \nu_F \mathbf{k}^{\prime} \cdot \sigma  }= \frac{(i\omega_n+\mu- \hbar v_F  p\mathbf{d} \cdot \mathbf{k}^{\prime})I_{2}+ \mathbf{k}^{\prime} \cdot \sigma}{(i\omega_n+\mu-\hbar v_F p\mathbf{d} \cdot \mathbf{k}^{\prime})^2-(\hbar v_F \mathbf{k}^{\prime})^2}
\end{equation}

\noindent For convenience, we set the constants $\hbar=v_F=1$ during the subsequent intermediate calculations. The relation between plasmons and the polarization function is provided by another important quantity that emerges in solid state physics, the dielectric function $\varepsilon(\nu,\mathbf{q})$, which is given by the relation 
\begin{equation}
    \varepsilon^R(\nu,\mathbf{q})= 1- V_0(\mathbf{q})\Pi^R(\nu,\mathbf{q})
\end{equation}

\noindent where $\Pi^R(\nu,\mathbf{q})$ is the retarded polarization function  and $V_0(\mathbf{q})=4\pi e^2/\kappa|\mathbf{q}|^2$ is the bare Coulomb potential with  dielectric constant $\kappa$. For simplicity, we will omit the superscript $R$ used for retarded quantities, i.e., we write for the retarded dielectric function $\varepsilon^R(\nu,\mathbf{q})=\varepsilon(\nu,\mathbf{q})$. This notation will also be used for other retarded quantities.\\

\noindent If we consider the inter-Weyl node scattering from the neighborhood of $\mathbf{k}_{W1}$ with the chemical potential $\mu_1$ to the neighborhood of $\mathbf{k}_{W2}$ with the chemical potential $\mu_2$, as discussed in the main text, the corresponding polarization operator $\Pi(i\nu,\mathbf{q})$ in the Matsubara frequency domain under the random phase approximation can be written as 
\begin{equation}
\label{eq:generequation}
    \Pi(i\nu,\mathbf{q})=\frac{1}{\beta} \sum_n \int \frac{d^3k}{(2\pi)^3} \text{Tr}\left[G(i\nu+i\omega_n+\mu_1,\mathbf{q}+\mathbf{k}_1,\mathbf{d})G(i\omega_n+\mu_2,\mathbf{k}_2,\mathbf{d})\right]  
\end{equation}

\noindent with $\mathbf{k}_1=\mathbf{k}-\mathbf{k}_{W1}$ and $\mathbf{k}_2=\mathbf{k}-\mathbf{k}_{W2}$, respectively.
   

\subsection{Derivation of the polarization function for Weyl semimetals}
\noindent To showcase the results at finite temperature, we return to Eq. (\ref{eq:generequation}). For simplicity we neglect the tilting factor from now on, which is already sufficient to explain the type-I Weyl semimetal behavior. We do the following transformations: $k_0=i\omega_n+\mu_2$, $q_0=i\nu+\mu_1-\mu_2$, $\mathbf{k}^{\prime}=\mathbf{k}-\mathbf{k}_{W2}$, and $\mathbf{q}^{\prime}=\mathbf{q}+\mathbf{k}_{W1}-\mathbf{k}_{W2}$. 
\begin{equation}
    \Pi(i\nu,\mathbf{q}) = \frac{2}{\beta} \sum_n \int \frac{d^3k^{\prime}}{(2\pi)^3}\frac{k_0(k_0+q_0)+\mathbf{k}^{\prime}\cdot (\mathbf{k}^{\prime}+\mathbf{q}^{\prime})}{\left((k_0+q_0)^2-E_{\mathbf{k}^{\prime}+\mathbf{q}^{\prime}}^2\right)\left(k_0^2-E_{\mathbf{k}^{\prime}}^2\right)}
\end{equation}

\noindent Afterwards, we perform the summation over Matsubara frequencies according to \cite{kapusta_2006}:
\begin{align*}
\begin{split}
    \frac{1}{\beta} \sum_{n} \frac{k_{0}\left(k_{0}+q_{0}\right)+\mathbf{k}^{\prime} \cdot(\mathbf{k}^{\prime}+\mathbf{q}^{\prime})}{(k_{0}^{2}-E_{\mathbf{k}^{\prime}}^{2})\left((k_{0}+q_{0})^2-E_{\mathbf{k}^{\prime}+\mathbf{q}^{\prime}}^{2}\right)} 
    =&-\frac{1}{2 \pi i} \oint dz\ h(z)g(z) \\
    =&-\frac{1}{2 \pi i} \left(\int\limits_{i \infty-\delta}^{-i \infty-\delta} dz\ h(z) \frac{1}{2} \tanh\left(\frac{\beta z}{2}\right)
       + \int\limits_{-i \infty+\delta}^{i \infty+\delta} dz\ h(z)\frac{1}{2}\tanh\left(\frac{\beta z}{2}\right)\right)\\
    =&-\frac{1}{2 \pi i} \left(\int\limits_{i \infty-\delta}^{-i \infty-\delta} dz\ h(z)\left(\frac{1}{2}-\frac{1}{e^{-\beta z}+1}\right)
        + \int\limits_{-i \infty+\delta}^{i \infty+\delta} dz\ h(z)\left(-\frac{1}{2}+\frac{1}{e^{\beta z}+1}\right)\right)
\end{split}
\end{align*}

\begin{align*}
    h(z)               &=\frac{(z+\mu_2)\left(z+\mu_2+q_{0}\right)+\mathbf{k}^{\prime} \cdot(\mathbf{k}^{\prime}+\mathbf{q}^{\prime})}{\left((z+\mu_2)^{2}-E_{\mathbf{k}^{\prime}}^2\right)\left(\left(z+\mu_2+q_{0}\right)^2-E_{\mathbf{k}^{\prime}+\mathbf{q}^{\prime}}^2\right)}\\
    g(z)               &=\frac{1}{2}\tanh\left(\frac{\beta z}{2}\right)\\ 
    E_{\mathbf{k}^{\prime}}    &=|\mathbf{k}^{\prime}|
\end{align*}
 
\noindent where the function $g(z)$ has simple poles at $i\omega_n$. We can divide the above expression into two parts in order to have two contributions for $\Pi(i\nu,\mathbf{q})$, the vacuum part and the matter part, such that 
\begin{align}
    \Pi_{\text{vac}}     &= -\frac{1}{2\pi i} \int\limits_{-i\infty}^{i\infty}dz\ \frac{1}{2}(h(z)+h(-z))\\
    \Pi_{\text{matter}}  &= \frac{1}{2\pi i} \int\limits_{-i \infty+\delta}^{i \infty+\delta} dz\ h(z) \frac{1}{e^{\beta z}+1}+\frac{1}{2\pi i} \int\limits_{i \infty-\delta}^{-i \infty-\delta} dz\ h(z) \frac{1}{e^{-\beta z}+1}
\end{align}

\noindent At this point in the derivation, we can simply evaluate the term corresponding to the vacuum part by using $z=i\omega$
\begin{align*}
\begin{split}
    \Pi_{\text{vac}}(i\nu,\mathbf{q})
    = &\int \frac{d\omega d^3k^{\prime}}{(2\pi)^4} \left(  \frac{(i\omega+\mu_2)(i\omega+\mu_2+q_0)+\mathbf{k}^{\prime}\cdot(\mathbf{k}^{\prime}+\mathbf{q}^{\prime})}{((i\omega+\mu_2+q_0)^2-|\mathbf{k}^{\prime}+\mathbf{q}^{\prime}|^2)((i\omega+\mu_2)^2-|\mathbf{k}^{\prime}|^2)}\right)\\
    & + \int \frac{d\omega d^3k^{\prime}}{(2\pi)^4} \left(\frac{(i\omega+\mu_2)(i\omega+\mu_2-q_0)+\mathbf{k}^{\prime}\cdot(\mathbf{k}^{\prime}+\mathbf{q}^{\prime})}{((i\omega+\mu_2-q_0)^2-|\mathbf{k}^{\prime}+\mathbf{q}^{\prime}|^2)((i\omega+\mu_2)^2-|\mathbf{k}^{\prime}|^2)}\right)\\
    \approx &-\frac{q^{\prime 2}}{24\pi^2} \left(\ln \left(-q_0^2+q^{\prime 2}+4\Lambda \right) -\ln (-q_0^2+q^{\prime 2}) +\frac{8\Lambda^3}{(-q_0^2+q^{\prime 2})^{3/2}}\text{arctan} \left(  \frac{\sqrt{-q_0^2+q^{\prime 2}}}{2\Lambda}\right) -\frac{4\Lambda^2}{-q_0^2+q^{\prime 2}}\right)
\end{split}
\end{align*}

\noindent where we relabeled $k^{\prime}=k$ again, set $q^{\prime}=|\mathbf{q}^{\prime}|$ and omitted the constant terms. We analytically extend the result above into real frequencies by $i\nu \to \nu +i\delta$:
\begin{equation}
\label{eq:polvac}
    \Pi_{\text{vac}}(\nu,\mathbf{q})=- \frac{q^{\prime 2}}{24\pi^2}\left(\ln \left( \frac{4\Lambda^2+q^{\prime 2}-\nu^{\prime 2}}{q^{\prime 2}-\nu^{\prime 2}}\right)+ \frac{8\Lambda^3}{(q^{\prime 2}-\nu^{\prime 2})^{3/2}} \text{arctan} \left(  \frac{\sqrt{q^{\prime 2}-\nu^{\prime 2}}}{2\Lambda}\right) -\frac{4\Lambda^2}{q^{\prime 2}-\nu^{\prime 2}} \right)
\end{equation}

\noindent with $\nu^{\prime}=\nu-\mu_1+\mu_2$. Notice that we have introduced a momentum cutoff $\Lambda$ since the integral over the momentum diverges. Afterwards, we can expand Eq. (\ref{eq:polvac}) in the case that $\Lambda\gg1$ to obtain:
\begin{equation}
    \Pi_{\text{vac}}(\nu,\mathbf{q}) \approx -\frac{|\mathbf{q}^{\prime}|^2}{24\pi} \left(\ln \left
   |\frac{4\Lambda^2}{|\mathbf{q}^{\prime}|^2-\nu^{\prime 2}}\right|+\frac{3(|\mathbf{q}^{\prime}|^2-\nu^{\prime 2})}{10\Lambda^2}-\frac{1}{3}\right)
\end{equation}

\noindent In order to obtain the real and imaginary parts of $\Pi_{vac}$, we can use the generalized Kramers-Kronig relation \cite{bjorken_1966}.
\begin{equation}
    \begin{split}
        \text{Re} [\Pi_{\text{vac}}(\nu,\mathbf{q})]= \text{Re} [\Pi_{\text{vac}}(0,\mathbf{q})] +\frac{E}{\pi}\mathcal{P} \int_{-\infty}^\infty d\omega \frac{\text{Im}[\Pi_{\text{vac}}(\nu,\mathbf{q})]}{\omega(\omega-E)}\\
        \text{Im} [\Pi_{\text{vac}}(\nu,\mathbf{q})]= \text{Im} [\Pi_{\text{vac}}(0,\mathbf{q})] -\frac{E}{\pi}\mathcal{P} \int_{-\infty}^\infty d\omega \frac{\text{Re}[\Pi_{\text{vac}}(\nu,\mathbf{q})]}{\omega(\omega-E)}
    \end{split}
\end{equation}

\noindent Therefore, $\Pi_\text{vac}$ is given by 
\begin{equation}
\label{eq:realimagpolvac}
    \Pi_{\text{vac}}(\nu,\mathbf{q})=-\frac{q^{\prime 2}}{24\pi^2} \left(\ln \left| \frac{4\Lambda^2+q^{\prime 2}-\nu^{\prime 2}}{q^{\prime 2}-\nu^{\prime 2}}\right|+ \frac{8\Lambda^3}{(q^{\prime 2}-\nu^{\prime 2})^{3/2}} \text{arctan} \left(  \frac{\sqrt{q^{\prime 2}-\nu^{\prime 2}}}{2\Lambda}\right) -\frac{4\Lambda^2}{q^{\prime 2}-\nu^{\prime 2}} + i\pi \Theta(\nu^{\prime}-q^{\prime})  \right)
\end{equation}

\noindent where we have considered $\text{Im}[\Pi_\text{vac}(0,\mathbf{q})]=0$. The imaginary part of $ \Pi_\text{vac}(\nu,\mathbf{q})$ is given by 
\begin{equation}
    \text{Im}[\Pi_{\text{vac}}(\nu,\mathbf{q})]=-\Theta(\nu^{\prime}-q^{\prime})\frac{q^{\prime 2}}{24\pi}
\end{equation}

\noindent The real part of Eq. (\ref{eq:realimagpolvac}) is given by
\begin{equation}
    \text{Re}[\Pi_{\text{vac}}(\nu,\mathbf{q})] = -\frac{q^{\prime 2}}{24\pi^2}\left(\ln \left| \frac{4\Lambda^2+q^{\prime 2}-\nu^{\prime 2}}{q^{\prime 2}-\nu^{\prime 2}}\right|+ \frac{8\Lambda^3}{(q^{\prime 2}-\nu^{\prime 2})^{3/2}} \text{arctan} \left(  \frac{\sqrt{q^{\prime 2}-\nu^{\prime 2}}}{2\Lambda}\right) -\frac{4\Lambda^2}{q^{\prime 2}-\nu^{\prime 2}} \right)
\end{equation}

\noindent We see that the logarithm always carries an imaginary part since its argument is negative. Let us evaluate the matter component of the polarization function. 
\begin{align*}
    \Pi_{\text{matter}}
    = &\frac{1}{2 \pi i} \int\limits_{-i \infty+\delta}^{i \infty+\delta} dz\ h(z) \frac{1}{e^{\beta z}+1}+\frac{1}{2 \pi i} \int\limits_{i \infty-\delta}^{-i \infty-\delta} dz\ h(z) \frac{1}{e^{-\beta z}+1}\\
    = &\ \frac{1}{2E_{\mathbf{k}^{\prime}}} \left(\frac{E_{\mathbf{k}^{\prime}}\left(E_{\mathbf{k}^{\prime}}+q_{0}\right)+\mathbf{k}^{\prime}\cdot(\mathbf{k}^{\prime}+\mathbf{q}^{\prime})}{\left(E_{\mathbf{k}^{\prime}}+q_{0}\right)^{2}-E_{\mathbf{k}^{\prime}+\mathbf{q}^{\prime}}^{2}}\right) \frac{1}{e^{\beta\left(E_{\mathbf{k}^{\prime}}+\mu_2\right)}+1}\\
        &+\frac{1}{2 E_{\mathbf{k}^{\prime}}} \left(\frac{E_{\mathbf{k}^{\prime}}\left(E_{\mathbf{k}^{\prime}}-q_{0}\right)+\mathbf{k}^{\prime}\cdot(\mathbf{k}^{\prime}+\mathbf{q}^{\prime})}{\left(E_{\mathbf{k}^{\prime}}-q_{0}\right)^{2}-E_{\mathbf{k}^{\prime}+\mathbf{q}^{\prime}}^{2}}\right) \frac{1}{e^{\beta\left(E_{\mathbf{k}^{\prime}}-\mu_2\right)}+1}\\
        &+\frac{1}{2E_{\mathbf{k}^{\prime}+\mathbf{q}^{\prime}}} \left(\frac{E_{\mathbf{k}^{\prime}+\mathbf{q}^{\prime}}\left(E_{\mathbf{k}^{\prime}+\mathbf{q}^{\prime}}+q_{0}\right)+\mathbf{k}^{\prime}\cdot(\mathbf{k}^{\prime}+\mathbf{q}^{\prime})}{\left(E_{\mathbf{k}^{\prime}+\mathbf{q}^{\prime}}+q_{0}\right)^{2}-E_{\mathbf{k}^{\prime}}^{2}}\right) \frac{1}{e^{\beta\left(E_{\mathbf{k}^{\prime}+\mathbf{q}^{\prime}}-\mu_1\right)}+1}\\
        &+\frac{1}{2 E_{\mathbf{k}^{\prime}+\mathbf{q}^{\prime}}} \left(\frac{E_{\mathbf{k}^{\prime}+\mathbf{q}^{\prime}}\left(E_{\mathbf{k}^{\prime}+\mathbf{q}^{\prime}}-q_{0}\right)+\mathbf{k}^{\prime}\cdot(\mathbf{k}^{\prime}+\mathbf{q}^{\prime})}{\left(E_{\mathbf{k}^{\prime}+\mathbf{q}^{\prime}}-q_{0}\right)^{2}-E_{\mathbf{k}^{\prime}}^{2}}\right) \frac{1}{e^{\beta\left(E_{\mathbf{k}^{\prime}+\mathbf{q}^{\prime}}+\mu_1\right)+1}}\\
    = &\ \frac{1}{2E_{\mathbf{k^{\prime}}}}\sum_{s=\pm}f^s\left(q_{0}\right) N_{F}^s\left(E_{\mathbf{k}^{\prime}}\right)
 \end{align*}
 
\begin{align}
    f^s\left(q_{0}\right) &=\frac{E_{\mathbf{k}'}\left(E_{\mathbf{k}'}+sq_{0}\right)+\mathbf{k}
    \cdot(\mathbf{k}'+\mathbf{q}')}{\left(E_{\mathbf{k}'}+sq_{0}\right)^{2}-E_{\mathbf{k}'+\mathbf{q}'}^{2}}\\
    N_{F}^{s}\left(E_{\mathbf{k}'}\right) &=\frac{1}{e^{\beta\left(E_{\mathbf{k}'}-s\mu_1\right)}+1}+\frac{1}{e^{\beta\left(E_{\mathbf{k}'}+s\mu_2\right)}+1}
\end{align}

\noindent where we considered the following change of variable: $\mathbf{k}^{\prime} \to -\mathbf{k}^{\prime}-\mathbf{q}^{\prime}$. If we take into account the finite density effect, the Fermi energy could lie either in the valence  band ($\mu_1$, $\mu_2<0$) or in the conduction band ($\mu_1$, $\mu_2>0$). 
\begin{align*}
    \Pi_{\text{matter}}(\nu,\mathbf{q},T)
    &= 2\int \frac{d^3k^{\prime}}{(2\pi)^3} \frac{1}{2E_{\mathbf{k}^{\prime}}}\sum_{s=\pm} f^s(q_0)N_{F}^s\left(E_{\mathbf{k}^{\prime}}\right)\\
    &= \frac{1}{8\pi^3}\int\limits_0^\infty dk^{\prime} \int\limits_0^{2\pi}d\phi \int\limits_0^\pi d\theta \sum_{s=\pm} k^2 \sin \theta \frac{s(\nu+\mu_1-\mu_2+i\delta)+2k^{\prime}+q^{\prime} \cos \theta}{(\nu+\mu_1-\mu_2+i\delta+s|k^{\prime}|)^2-(k^{\prime 2}+q^{\prime 2}+2k^{\prime}q^{\prime}\cos \theta)}N_F^s(|k^{\prime}|)
\end{align*} 

\noindent We can obtain the real part of $\Pi_{\text{matter}}(\nu,\mathbf{q},T)$
\begin{equation}
    \text{Re}[\Pi_{\text{matter}}(\nu,\mathbf{q},T)]=\frac{1}{8\pi^3}\int_0^{\infty} dk' \sum_{s=\pm} J^{s}(\nu,\mathbf{k}',\mathbf{q}',T)
\end{equation}
\begin{equation}
    J^{s}(\nu,\mathbf{k}^{\prime},\mathbf{q}^{\prime},T)= -2\pi k^{\prime}N_F^s(|k^{\prime}|)-\pi \frac{(2s k^{\prime}+\nu^{\prime})^2-q^{\prime 2}}{2q^{\prime}}N_F^s(|k^{\prime}|)\ln \left |  \frac{(q^{\prime}-s \nu')(2k^{\prime}+s\nu+q^{\prime})}{(s\nu^{\prime}+q')(q^{\prime}-2k^{\prime}-s\nu^{\prime})}\right|
\end{equation}

\noindent At nonzero temperatures, the momentum space integral cannot be evaluated analytically. Therefore, we initially compute the polarization function at zero temperature in which an analytic treatment is adequate. The zero temperature scenario, despite being simple, still exhibits the Kohn anomaly observed in  experiment. 

\subsection{Zero temperature results}
\noindent At zero temperature, $N_F^s(E_{\mathbf{k}^{\prime}})$ can be replaced by the Heaviside step functions as the following 
\begin{equation}
    \lim_{\beta \to \infty }N_F^s(E_{\mathbf{k}^{\prime}})=\Theta(s\mu_1-E_{\mathbf{k^{\prime}}})+\Theta(-s\mu_2-E_{\mathbf{k^{\prime}}})
\end{equation}

 \noindent Hereafter, we consider only the absolute values of the chemical potentials $|\mu_1|$ and $|\mu_2|$ and we only present results for $\mu_1$, $\mu_2> 0$. Thus, at zero temperature, we obtain
\begin{align*}
\textrm{Re}[\Pi_\text{matter}(\nu,\mathbf{q},T=0)]=
&\ -\frac{1}{6\pi^2}(\mu_{1}^{2}+\mu_{2}^{2})-\frac{1}{24\pi^2}\nu^{\prime}(\mu_{1}-\mu_{2})\\
&+\frac{q^{\prime 2}}{32\pi^2}\sum_{s=\pm1}g_{3}\left(\frac{2\mu_{1}+\nu^{\prime}}{sq^{\prime}}\right)
\ln\left|\frac{2\mu_{1}+\nu^{\prime}-sq'}{\nu^{\prime}-sq'}\right|\\
&+\frac{q^{\prime 2}}{32\pi^2}\sum_{s=\pm1}g_{3}\left(\frac{2\mu_{2}-\nu^{\prime}}{sq^{\prime}}\right)\ln \left|\frac{2\mu_{2}-\nu^{\prime}-sq^{\prime}}{\nu^{\prime}+sq^{\prime}}\right|
\end{align*}

\noindent where
\begin{equation*}
    g_{3}(x)=x\left(\frac{x^{2}}{3}-1\right)+\frac{2}{3} \qquad \nu^{\prime}=\nu-\mu_1+\mu_2
\end{equation*}

\noindent The zero-temperature results are consistent with recent calculations \cite{lv_2013,zhou_2015,thakur_2018,sadhukhan_2020}. Adding the vacuum and the matter parts together, we obtain the real part of the polarization function at zero temperature
\begin{equation}\label{box}
  \begin{split}\operatorname{Re}[\Pi(\nu,\mathbf{q}, T=0)]
  =& \frac{q^{\prime 2}}{32\pi^2}\sum_{s=\pm1}g_{3}\left(\frac{2\mu_{1}+\nu^{\prime}}{sq^{\prime}}\right)\ln\left|\frac{2\mu_{1}+\nu^{\prime}-sq^{\prime}}{\nu^{\prime}-sq^{\prime}}\right| +\frac{q^{\prime 2}}{32\pi^2}\sum_{s=\pm1}g_{3}\left(\frac{2\mu_{2}-\nu^{\prime}}{sq^{\prime}}\right)\ln \left|\frac{2\mu_{2}-\nu^{\prime}-sq^{\prime}}{\nu^{\prime}+sq^{\prime}}\right|\\
& -\frac{1}{6\pi^2}(\mu_{1}^{2}+\mu_{2}^{2})-\frac{1}{24\pi^2}(\nu^{\prime} \mu_{1}-\nu^{\prime} \mu_{2}) -\frac{q^{\prime 2}}{24 \pi^2}\left(\ln\left|\frac{4\Lambda^2}{q^{\prime 2}-\nu^{\prime 2}}\right|
-\frac{1}{3}+\frac{3(q^{\prime 2}-\nu^{\prime 2})}{10\Lambda^2}\right)
\end{split}
\end{equation}

\noindent Now, we calculate the imaginary part of the polarization function for the matter term  using $\Theta=\frac{ \mathbf{k^{\prime}}\cdot (\mathbf{k^{\prime}+q^{\prime}})}{|\mathbf{k^{\prime}}||\mathbf{k^{\prime}+q^{\prime}}|}$ to obtain 
\begin{align*}
 \text{Im}[\Pi_\text{matter}(\nu,\mathbf{q},T)]
 =&\int \frac{d^3k^{\prime}}{(2\pi)^3}\frac{(\nu+\mu_1-\mu_2+i\delta)+|\mathbf{k^{\prime}}|+|\mathbf{k^{\prime}+q^{\prime}}|\cos \Theta}{(\nu+\mu_1-\mu_2+i\delta +|\mathbf{k^{\prime}}|)^2-|\mathbf{k^{\prime}+q^{\prime}}|^2}N_F^{+}(|\mathbf{k^{\prime}}|)\\
 &-\int \frac{d^3k^{\prime}}{(2\pi)^3}\frac{(\nu+\mu_1-\mu_2+i\delta)-|\mathbf{k^{\prime}}|-|\mathbf{k^{\prime}+q^{\prime}}|\cos \Theta}{(\nu+\mu_1-\mu_2+i\delta -|\mathbf{k^{\prime}}|)^2-|\mathbf{k^{\prime}+q^{\prime}}|^2}N_F^{-}(|\mathbf{k^{\prime}}|)\\
 =&\ \text{Im}[\Pi_\text{matter}^{+}(\nu,\mathbf{q},T)]+\text{Im}[\Pi_\text{matter}^{-}(\nu,\mathbf{q},T)]
 \label{imaginarypart}
\end{align*}

\noindent with the following terms 

\begin{align*}
\text{Im}[\Pi_\text{matter}^{+}(\nu,\mathbf{q},T)]=&\ -\frac{\Theta(q^{\prime}-\nu^{\prime})}{16\pi} \Theta(q^{\prime}+\nu^{\prime}) \int\limits_{\frac{q^{\prime}-\nu^{\prime}}{2}}^{\infty}dk^{\prime}N_F^{+}(k^{\prime})\frac{(2k^{\prime}+\nu')^2-q^{\prime 2}}{q^{\prime}}\\
     &-\frac{\Theta(-q^{\prime}-\nu^{\prime})}{16\pi}\Theta(q^{\prime}-\nu^{\prime})\int_0^\infty dk^{\prime} N_F^{+}(k^{\prime})\frac{(2k^{\prime}+\nu^{\prime})^2-q^{\prime 2}}{q'} \left [\Theta \left(\frac{q^{\prime}-\nu^{\prime}}{2}-k^{\prime}  \right) - \Theta \left ( \frac{-\nu^{\prime}-q^{\prime}}{2}-k^{\prime}\right)\right]
\end{align*}

\begin{align*}
   \text{Im}[\Pi_\text{matter}^{-}(\nu,\mathbf{q},T)]= &+\frac{\Theta(q^{\prime}+\nu^{\prime})}{16\pi} \Theta(q^{\prime}-\nu^{\prime}) \int\limits_{\frac{q^{\prime}+\nu^{\prime}}{2}}^{\infty}dk^{\prime}N_F^{-}(k^{\prime}) \frac{(-2k^{\prime}+\nu^{\prime})^2-q^{\prime 2}}{q^{\prime}}\\  &-\frac{\Theta(q^{\prime}+\nu^{\prime})}{16\pi}\Theta(\nu^{\prime}-q^{\prime})\int_{0}^{\infty}dk^{\prime}N_F^{-}(k^{\prime})\frac{(-2k^{\prime}+\nu^{\prime})^2-q^{\prime 2}}{q^{\prime}} \left[\Theta\left(\frac{q^{\prime}+\nu^{\prime}}{2}-k^{\prime}  \right)-\Theta \left(\frac{\nu^{\prime}-q^{\prime}}{2} -k^{\prime} \right)  \right]
\end{align*}

\noindent Combining the vacuum and matter parts, we obtain the imaginary part of $\Pi(\nu,\mathbf{q}, T=0)$ at zero temperature as follows 

 \begin{equation}
 \begin{split}\text{Im} [\Pi (\nu,\mathbf{q}, T=0)]=\enskip& -\frac{q^{\prime 2}}{24\pi}\Theta(\nu^{\prime}-q^{\prime})\\
 &-\frac{q^{\prime 2}}{32 \pi}\Theta (q^{\prime}-\nu^{\prime}) \Theta(q^{\prime}+\nu^{\prime})\Theta\left(\frac{2 \mu_1+\nu^{\prime}}{q^{\prime}}-1\right) g_{3}\left(\frac{2 \mu_1+ \nu^{\prime}}{q^{\prime}}\right)\\
 &-\frac{q^{\prime 2}}{32\pi}\Theta(-q^{\prime}-\nu^{\prime})\Theta(q^{\prime}-\nu^{\prime})\Theta \left(  1-\frac{2\mu_1+\nu^{\prime}}{q^{\prime}}\right)g_3 \left(\frac{2\mu_1+\nu^{\prime}}{q^{\prime}} \right)\\
    &+\frac{q^{\prime 2}}{32\pi}\Theta(-q^{\prime}-\nu^{\prime}) \Theta(q^{\prime}-\nu^{\prime})\Theta \left(  -1-\frac{2\mu_1+\nu^{\prime}}{q^{\prime}}\right)g_3 \left(\frac{2\mu_1+\nu^{\prime}}{q^{\prime}} \right)\\
    &+\frac{q^{\prime 2}}{24\pi}\Theta(-q^{\prime}-\nu^{\prime})\Theta(q^{\prime}-\nu^{\prime})\Theta \left( \frac{2\mu_1+\nu^{\prime}}{q^{\prime}}+1\right)\\
 &+\frac{q^{\prime 2}}{32\pi}\Theta(q^{\prime}+\nu^{\prime})\Theta(q^{\prime}-\nu^{\prime})\Theta \left(\frac{2\mu_2-\nu^{\prime}}{q^{\prime}}-1 \right)g_3\left( \frac{2\mu_2-\nu^{\prime}}{q^{\prime}}  \right)\\
     &-\frac{q^{\prime 2}}{32\pi}\Theta(\nu^{\prime}-q^{\prime})\Theta(q^{\prime}+\nu^{\prime})\Theta\left(1-\frac{2\mu_2-\nu^{\prime}}{q^{\prime}} \right)g_3\left( \frac{2\mu_2-\nu^{\prime}}{q^{\prime}}  \right)\\
     &+\frac{q^{\prime 2}}{32\pi}\Theta(\nu^{\prime}-q^{\prime})\Theta(q^{\prime}+\nu^{\prime})\Theta\left(-1-\frac{2\mu_2-\nu^{\prime}}{q^{\prime}} \right)g_3\left( \frac{2\mu_2-\nu^{\prime}}{q^{\prime}}  \right)\\
          &+\frac{q^{\prime 2}}{24\pi}\Theta(\nu^{\prime}-q^{\prime})\Theta(q^{\prime}+\nu^{\prime})\Theta \left(  \frac{2\mu_2-\nu^{\prime}}{q^{\prime}}+1\right)
\end{split}
\end{equation}

\subsection{Kohn anomaly from singularities of the polarization function \texorpdfstring{$\Pi(\nu,\textbf{q})$}{P}}

\noindent The Kohn anomaly is an anomaly in the dispersion relation and describes a sudden dip in frequency for a particular wavevector. The phonon dispersion relation is given by
\begin{equation}\label{Kohn}
\omega_{\mathbf{q}}^2=\frac{\Omega_{\mathbf{q}}^2}{\varepsilon(\nu,\mathbf{q})} \qquad \varepsilon (\nu,\mathbf{q})=1- V_0(\mathbf{q}) \text{Re}[\Pi(\nu,\mathbf{q})]
\end{equation} 

\noindent where $\omega_{\mathbf{q}}$ and $\Omega_{\mathbf{q}}$ are the renormalized phonon frequency due to electronic RPA screening and the bare phonon frequency, respectively. The subscript $r$ means the designates part. Thus, in order to find the Kohn anomaly, we must study the singularities of  $\Pi(\nu,\mathbf{q})$.\\

\noindent One might naively think that $\operatorname{Re}[\Pi(\nu, \mathbf{q})]$ has poles at $q^{\prime}=|\nu^{\prime}|$, $q^{\prime}=|2\mu_1 + \nu^{\prime}|$ and $q^{\prime}=|2\mu_2 - \nu^{\prime}|$. However, upon closer inspection, this guess turns out to be only partly correct. In the case $q^{\prime}=|2\mu_1 + \nu^{\prime}|$ , observe from Eq. (\ref{box}) that the apparently divergent term is of the form 
\begin{equation}
   x \to 0 : g_{3}(1+x)\ln{|x|} \sim x^2\ln{|x|} = 0 \qquad x\equiv \frac{|2\mu_1 + \nu^{\prime}|}{q^{\prime}}-1  
\end{equation}

\noindent and therefore vanishes. This is consistent with the fact that $\operatorname{Im}[\Pi(\nu, \mathbf{q})]$ is continuous and well defined  at $q^{\prime}=|2\mu_1 + \nu^{\prime}|$. Similarly, the alleged pole at $q^{\prime}=|2\mu_2 - \nu^{\prime}|$ disappears upon closer scrutiny.\\

\noindent For the case $q^{\prime} \to |\nu^{\prime}|$, by taking the most divergent terms, we see from Eq. (\ref{box}) that 
\begin{equation}\label{realPidiv}
 \operatorname{Re}[ \Pi(\nu, \mathbf{q})] \simeq \frac{q^{\prime 2}}{32 \pi^2}\left( g_3\left(1+\frac{2\mu_1}{\nu^{\prime}}\right)\ln\left|\frac{2\mu_{1}}{q^{\prime}-|\nu^{\prime}|}\right|+g_3\left(1-\frac{2\mu_2}{\nu^{\prime}}\right)\ln\left|\frac{2\mu_{2}}{q^{\prime}-|\nu^{\prime}|}\right|\right)-\frac{q^{\prime 2}}{24 \pi^2}\ln{\left|\frac{4\Lambda^2}{q^{\prime 2}-\nu^{\prime 2}}\right|}
 \to \infty
\end{equation}
\noindent This divergence of $\operatorname{Re}[\Pi(\nu, \mathbf{q})]$ goes hand in hand with the discontinuity of $\operatorname{Im}[\Pi(\nu, \mathbf{q})]$ at $q^{\prime} = |\nu^{\prime}|$. 

\begin{equation}
\begin{aligned}
    \frac{\partial}{\partial q^{\prime}}\operatorname{Re}[ \Pi(\nu, \mathbf{q})] \simeq &\frac{q^{\prime}}{16 \pi^2}\left( g_3\left(1+\frac{2\mu_1}{\nu^{\prime}}\right)\ln\left|\frac{2\mu_{1}}{q^{\prime}-|\nu^{\prime}|}\right|+g_3\left(1-\frac{2\mu_2}{\nu^{\prime}}\right)\ln\left|\frac{2\mu_{2}}{q^{\prime}-|\nu^{\prime}|}\right|\right)
    -\frac{q^{\prime}}{12 \pi^2}\ln{\left|\frac{4\Lambda^2}{q^{\prime 2}-\nu^{\prime 2}}\right|}\\
    &-\frac{q^{\prime 2}}{32\pi^2}\left(g_3\left(1+\frac{2\mu_1}{\nu^{\prime}}\right)\frac{1}{q^{\prime}-|\nu^{\prime}|}+g_3\left(1-\frac{2\mu_2}{\nu^{\prime}}\right)\frac{1}{q^{\prime}-|\nu^{\prime}|}\right)+\frac{1}{12\pi^2}\frac{q^{\prime 3}}{q^{\prime 2}-\nu^{\prime 2}}
\end{aligned}    
\end{equation}

\noindent The derivative of the real part of the dielectric function is

\begin{equation}
    \frac{\partial \varepsilon}{\partial q}=-2V_{0}(\mathbf{q})\text{Re}[\Pi(\nu,\mathbf{q})]\left(\frac{1}{q^{\prime}}-\frac{1}{q}\right)-V_{0}(\mathbf{q})\frac{\partial}{\partial q^{\prime}} \text{Re}[\Pi(\nu, \mathbf{q})]
\end{equation}

\noindent The first term is logarithmically divergent from Eq. (\ref{realPidiv}) and the second term includes both the logarithmic and power-law divergence. By only including the power-law divergence, which diverges faster than the logarithmic divergence, we find in the limit $q^{\prime} \rightarrow |\nu^{\prime}|$

\begin{equation}
    \frac{\partial \varepsilon}{\partial q}\propto -V_{0}(\mathbf{q})\left(\frac{q^{\prime 2}}{32\pi^2}\left(g_3\left(1+\frac{2\mu_1}{\nu^{\prime}}\right)\frac{1}{q^{\prime}-|\nu^{\prime}|}+g_3\left(1-\frac{2\mu_2}{\nu^{\prime}}\right)\frac{1}{q^{\prime}-|\nu^{\prime}|}\right)+\frac{1}{12\pi^2}\frac{q^{\prime 3}}{q^{\prime 2}-\nu^{\prime 2}}\right)
\end{equation}

\noindent Therefore, we observe a Kohn anomaly at $q^{\prime} = |\nu^{\prime}|=|\nu-\mu_1+\mu_2|$, where the polarization function blows up.  Notice that the location of the pole only depends on the difference between the chemical potentials of the two nodes.

\subsection{Finite temperature results}
\noindent In the following section, we take into account the finite temperature-dependence of the polarization. The real part of the polarization function at $T>0$ is 
 \begin{equation}
 \begin{split}
    \text{Re}[\Pi(\nu,\mathbf{q},T)]= &-\frac{q^{\prime 2}}{24\pi^2}\left( \ln \left| \frac{4\Lambda^2}{q^{\prime 2}-\nu^{\prime 2}}\right|+ \frac{3(q^{\prime 2}-\nu^{\prime 2})-\frac{1}{3}}{10\Lambda^2} \right)+\frac{\Gamma(2)}{4\pi^2}\sum_{s=\pm} \frac{1}{\beta^2}\left [\text{Li}_{2}\left(-e^{s\beta \mu_1}\right)+ \text{Li}_{2}\left(-e^{-s\beta \mu_2}\right)\right]\\
    &-\frac{1}{32\pi^2} \left(\frac{\nu^{\prime}}{q^{\prime}}F_{+}(\nu,q,T)+ \frac{\nu'}{q^{\prime}}F_{-}(\nu,q,T) \right)
\end{split}
\end{equation}

\noindent with 

\begin{equation}
    F_{\pm}(\nu,q,T)= \int\limits_{\pm 1}^{\infty}du \left(\frac{(\nu^{\prime} u)^2-(q^{\prime})^2}{\text{exp}\left(\frac{(\nu^{\prime} u\mp \nu^{\prime})\pm2 \mu_2}{2k_BT}\right)+1}+ \frac{(\nu^{\prime} u)^2-(q^{\prime})^2}{\text{exp}\left(\frac{(\nu^{\prime} u\mp \nu^{\prime})\mp2 \mu_1}{2k_BT}\right)+1}\right)\ln \left|\frac{(\pm\nu^{\prime}-q^{\prime})(\nu^{\prime}u+q^{\prime})}{(\pm \nu^{\prime}+q^{\prime})(\nu^{\prime}u-q^{\prime})}\right|
\end{equation}

\noindent The imaginary part of the polarization function at finite temperatures $T>0$ can be expressed as 
 \begin{equation}
 \begin{split}
     \text{Im}[\Pi(\nu,\mathbf{q},T)]=&\ -\frac{q^{\prime 2}}{32\pi}\Theta(q^{\prime}+\nu')\left(\Theta(q^{\prime}-\nu')\left(G_{+}(\nu,\mathbf{q},T)+G_{-}(\nu,\mathbf{q},T)\right)- \Theta(\nu^{\prime}-q^{\prime})H(\nu,\mathbf{q},T) \right)\\
     &-\frac{q^{\prime 2}}{24\pi}\Theta(\nu^{\prime}-q^{\prime})
     - \frac{q^{\prime 2}}{32\pi^2}\Theta(-q^{\prime}-\nu')\Theta(q^{\prime}-\nu^{\prime})M(\nu,q,T)
    \end{split}
\end{equation}

\begin{equation}
    G_{\pm}(\nu,\mathbf{q},T)= \int\limits_1^{\infty}du\left( \frac{u^2-1}{\text{exp}\left(\frac{(q^{\prime}u \mp \nu^{\prime})\pm2 \mu_2}{2k_BT}\right)+1}+\frac{u^2-1}{\text{exp}\left(\frac{(q^{\prime}u \mp \nu^{\prime})\mp 2\mu_1}{2k_BT}\right)+1}\right)
\end{equation}
\begin{equation}
    H(\nu,\mathbf{q},T)=\int\limits_{-1}^{1}du \left( \frac{u^2-1}{\text{exp}\left(\frac{(q^{\prime}u + \nu^{\prime})-2 \mu_2}{2k_BT}\right)+1}+\frac{u^2-1}{\text{exp}\left(\frac{(q^{\prime}u + \nu^{\prime})+2 \mu_1}{2k_BT}\right)+1}\right)
\end{equation}
\begin{equation}
    M(\nu,q,T)=\int\limits_{-1}^{1}du \left( \frac{u^2-1}{\text{exp}\left(\frac{(q^{\prime}u - \nu^{\prime})+2 \mu_2}{2k_BT}\right)+1}+\frac{u^2-1}{\text{exp}\left(\frac{(q^{\prime}u - \nu^{\prime})-2 \mu_1}{2k_BT}\right)+1}\right)
\end{equation}

\noindent While for finite temperatures, exact analytic expressions cannot easily be obtained, it is possible to obtain numerical results from our semi-analytical results.

\subsection*{Supplementary B: Single crystal growth and sample preparation of TaP for IXS}
\label{sec:supplementaryB}
\noindent Single crystals of tantalum phosphide (TaP) were prepared using the chemical vapor transport method. 3g of Ta (Beantown Chemical, 99.95\%) and P (Beantown Chemical, 99.999\%) powders were weighed and mixed together inside a glovebox. They were subsequently flame-sealed inside a quartz tube and then heated to 70$^\circ$C to be held for 20 hours before a pre-reaction. Afterwards, the obtained TaP powder was sealed inside another quartz tube with 0.4g of I$_2$ (Sigma Aldrich, $\geq$ 99.8\%) and this tube was horizontally placed in a two-zone furnace. To improve the crystal size and quality, instead of setting a 100$^\circ$C temperature difference, we gradually increased the temperature difference from zero until the point the I$_2$ transport agent started to flow. This process seems to be furnace- and distance-specific. In our case, the optimal temperatures for the two zones were 900$^\circ$C and 950$^\circ$C, respectively, and the distance between the two heating zones was constantly optimized. With the help of the transport agent I$_2$, the TaP source materials were transferred from the cold end of the tube to the hot end and condensed into single-crystalline form within 14 days. The single crystals are centimeter-sized and have a metallic luster (Figure \ref{fig:sup3}). The Fermi level information of this sample is well-characterized in a separate study \cite{han_2019}. 

Due to the high X-ray absorption of tantalum and the large c-axis dimension of the crystals grown in laboratory, the thickness of the samples were required to be reduced to a suitable value for performing inelastic x-ray scattering (IXS) experiments. We determined that the optimal sample thickness for our experiment configuration corresponding to an X-ray wavelength of 0.5725\text{\AA} was $\sim$20$\mu$m, allowing for a photon transmission of 0.33. A portion of the crystal had its orientation determined using a back-scattering Laue diffractometer and afterwards, it was thinned down to $\sim$20$\mu$m by polishing followed by being glued onto a brass sample holder with a GE-vanish (as the other end of the sample holder was connected to the cryostat). Figure \ref{fig:sup1} shows top and side views of the orientated thinned-down sample used for our IXS experiments.

\subsection*{Supplementary C: X-ray and neutron scattering measurements and analysis of TaP}
\label{sec:supplementaryC}

\textit{Inelastic X-ray scattering experiments.} Inelastic X-ray scattering measurements were performed on the high-energy resolution inelastic X-ray (HERIX) instrument at sector 3-ID beamline of the Advanced Photon Source, Argonne National Laboratory, with incident beam energy of 21.657keV ($\lambda=0.5725$\text{\AA}) and overall energy resolution of 2.1meV \cite{sinn_2001,alatas_2011,toellner_2011}. The incident beam was focused on the sample using a toroidal and KB mirror system. The full width at half maximum (FWHM) of the beam size at sample position was 20x20$\mu$m$^2$ (V$\times$H). The spectrometer was functioning in the horizontal scattering geometry with horizontally polarized radiation. The scattered beam was analyzed by diced and spherically curved silicon (18 6 0) analyzers working at the backscattering angle. The measurements were performed at temperatures of 18K, 60K, 100K and 300K. \\

\noindent \textit{Inelastic neutron scattering experiments.} Inelastic neutron scattering (INS) was performed at the HB1 polarized triple-axis spectrometer at the High-Flux Isotope Reactor at the Oak Ridge National Laboratory. We used a fixed $E_f = 14.7$meV with 48$^\prime - $40$^\prime - $40$^\prime - $120$^\prime$ collimation and Pyrolytic Graphite filters to eliminate higher-harmonic neutrons. Measurements were performed using closed-cycle refrigerators between room temperature and the base temperature of 4K. INS measurements were also performed at the BT-7 double focusing triple-axis spectrometer \cite{lynn_2012} at the NIST Center for Neutron Research at the National Institute of Standards and Technology. At this research facility, we used the same fixed final energy with an open$ -$80$^\prime - $80$^\prime - $120$^\prime$ collimation and Pyrolytic Graphite filters. Measurements were performed at 10K.\\

\noindent \textit{Analysis of IXS and INS experimental data.} Statistical error is taken to be square root of the number of counts from Poisson statistics. Repeated IXS scans were performed for certain $\boldsymbol{q}$ points and subsequently merged together to reduce statistical noise. Spectra obtained from constant wavevector IXS scans (measuring intensity of counts versus energy transfer) were normalized by area and fitted using a fitting core function of damped harmonic oscillators that were convoluted with a pseudo-Voigt function representing the instrument resolution function, measured before the experiment took place. The number of Lorentzian peaks, corresponding to the number of phonon modes, added to the fitting core was known from \textit{ab initio} phonon dispersion calculations within the measured energy range. Each fitted peak has parameters corresponding to center, FWHM and amplitude. The latter was furthermore modulated during the fit to take into account the longitudinal and transversal modes of propagation with respect to the scan direction. In our case, due to the inadequate energy resolution necessary for a precise FWHM measurement, more care was taken into the fit to extract accurate peak center locations, corresponding to the energy of the phonon mode. Plots of the phonon dispersion shown in Figures \hyperref[fig:2]{2f} and \ref{fig:sup6} are created by plotting the fitted core after deconvolution with the instrumental resolution function. The INS scans were treated in an analogous manner by using the Data Analysis and Visualization Environment (\textlcsc{DAVE}) software \cite{dave_2009} and \textlcsc{NeutronPy} based on the ResLib program package. The resolution function was calculated with knowledge of the monochromator and analyzer crystals, the collimation as well as the sample configuration of our experiment in \textlcsc{NeutronPy}. The phonon modes were modeled using Lorentzian functions as was done for IXS. For the INS data, the intensity near the elastic peak was difficult to resolve for phonon modes located near this energy transfer due to poorer energy resolution and were therefore neglected in favor of extracting phonon modes with larger energy transfer with these INS data scans.\\

\noindent \textit{X-ray and neutron diffraction}. X-ray and neutron diffraction measurements of single-crystalline TaP were performed prior to performing the inelastic scattering measurements. These were required for the calculation of the orientation matrix used for alignment purposes and for the determination of the crystal lattice parameters at different temperatures. A couple of these scans near the elastic Bragg peaks at (004) and (200) obtained from neutron scattering are shown in Figure \ref{fig:sup10}. Values obtained from both x-ray and neutron diffraction measurements at room temperature agree well with those obtained in Ref. \cite{xu_2015} ($a = b = 3.32$\text{\AA} and $c = 11.34$\text{\AA}). 

\subsection*{Supplementary D: Computational details}

\noindent All \textit{ab initio} calculations are performed using the VASP \cite{kresse_1999, kresse_1996, kresse_1995} with projector-augmented-wave (PAW) pseudopotentials and Perdew-Burke-Ernzerhof (PBE) for exchange-correlation energy functional \cite{perdew_1996}. The geometry optimization of the conventional cell was performed with a 6x6x2 Monkhorst-Pack grid of k-point sampling. The second-order and third-order force constants were calculated using a real space supercell approach with a 3x3x1 supercell. The Phonopy package was used to obtain the second-order force constants used in the calculation \cite{togo_2015}.

\subsection*{Supplementary E: Phonon nesting conditions between Weyl nodes in TaP}
\label{sec:supplementaryD}
\noindent In Table S1, we list the possible phonon nesting conditions for a phonon $\mathbf{q} = \mathbf{k}_{Wj_1} - \mathbf{k}_{Wj_2}$ between two Weyl electronic states at $\mathbf{k}_{Wj_1}$ and $\mathbf{k}_{Wj_2}$. Furthermore, we present the momentum-mismatch between phonon $\mathbf{q}$ and the Weyl node location in momentum-space $\mathbf{k}_{Wj'}$ (which is reduced by the dynamical effect described in the main text) as well as the chirality of the two Weyl electronic states to highlight their conservation within the context of the topological Kohn anomaly in this material. The momentum-space coordinates of the 8 W1 ($k_z=0$) and 16 W2 ($k_z\neq0$) Weyl nodes of TaP along with their respective chiralities are known from \cite{weng_2015}.

\begin{table}[ht!]
\label{table:nesting-conditions}
\caption{Possible phonon nesting conditions between different Weyl electronic states within TaP. Along with the momentum-space coordinates of the Weyl node in the Brillouin zone, the type and chirality of the electronic state is shown. Momentum conservation (presented in the form of a mismatch between nesting phonon $\mathbf{q} = \mathbf{k}_1 - \mathbf{k}_2$ and the Weyl node location $\mathbf{k}_{Wj^{\prime}}$, i.e. |$\mathbf{q}-\mathbf{k}_{Wj}|/|\mathbf{k}_{Wj}|$ or $|\mathbf{q}|$ in the case of $\Gamma$) and the chirality conservation are featured in the table. Only one representative set for each case between Weyl nodes is shown with the number of instances indicated.}
\resizebox{\textwidth}{!}{
\begin{tabular}{c c c?c c c?c?c?c?c}
& & & & & & Phonon & Momentum & Chirality &  \\
$\mathbf{k}_1$ & Type & Chirality & $\mathbf{k}_2$ & Type & Chirality & $\mathbf{q}$ & conservation & conservation & Multiplicity\\
\hline
($+$0.271, $-$0.024, $+$0.578)  & W2    & + & ($+$0.518, $+$0.014, $+$0.000)   & W1    & + & W2 & 4.35\%    & \checkmark    & 16\\
($+$0.271, $+$0.024, $+$0.578)   & W2    & - & ($+$0.518, $+$0.014, $+$0.000) & W1    & + & W2 & 4.35\%    & $\times$ & 16\\
($-$0.271, $+$0.024, $+$0.578)  & W2    & + & ($+$0.518, $+$0.014, $+$0.000)   & W1    & + & W2 & 4.49\%    & \checkmark    & 16\\
($-$0.271, $-$0.024, $+$0.578)   & W2    & - & ($+$0.518, $+$0.014, $+$0.000) & W1    & + & W2 & 4.49\%    & $\times$ & 16\\
($+$0.024, $+$0.271, $+$0.578)   & W2    & + & ($+$0.518, $+$0.014, $+$0.000)  & W1    & +     & W2 & 34.98\%   & \checkmark    & 16\\
($+$0.024, $-$0.271, $+$0.578)  & W2    & - & ($+$0.518, $+$0.014, $+$0.000) & W1    & + & W2 & 34.98\%    & $\times$ & 16\\
($-$0.024, $-$0.271, $-$0.578)   & W2    & + & ($+$0.518, $+$0.014, $+$0.000)  & W1    & +     & W2 & 42.48\%   & \checkmark    & 16\\
($-$0.024, $+$0.271, $+$0.578)  & W2    & - & ($+$0.518, $+$0.014, $+$0.000) & W1    & + & W2 & 42.48\%    & $\times$ & 16\\
\hline
($+$0.271, $+$0.024, $+$0.578)  & W2    & - & ($-$0.271, $+$0.024, $+$0.578)   & W2    & + & W1 & 5.36\%    & $\times$    & 16\\
($+$0.271, $+$0.024, $+$0.578)  & W2    & - & ($+$0.271, $-$0.024, $+$0.578)   & W2    & + & W1 & 7.10\%    & $\times$    & 16\\
($+$0.271, $+$0.024, $+$0.578)  & W2    & - & ($-$0.271, $-$0.024, $+$0.578)   & W2    & - & W1 & 8.03\%    & \checkmark    & 16\\
($+$0.271, $+$0.024, $+$0.578)  & W2    & - & ($+$0.271, $+$0.024, $-$0.578)   & W2    & - & W1 & 30.35\%    & \checkmark    & 16\\
($+$0.271, $+$0.024, $+$0.578)  & W2    & - & ($-$0.271, $+$0.024, $-$0.578)   & W2    & + & W1 & 30.58\%    & $\times$    & 16\\
($+$0.271, $+$0.024, $+$0.578)  & W2    & - & ($+$0.271, $-$0.024, $-$0.578)   & W2    & + & W1 & 30.93\%    & $\times$    & 16\\
($+$0.271, $+$0.024, $+$0.578)  & W2    & - & ($-$0.271, $-$0.024, $-$0.578)   & W2    & - & W1 & 31.16\%    & \checkmark    & 16\\
($+$0.271, $+$0.024, $+$0.578)  & W2    & - & ($-$0.024, $-$0.271, $+$0.578)   & W2    & + & W1 & 60.86\%    & $\times$    & 16\\
($+$0.271, $+$0.024, $+$0.578)  & W2    & - & ($-$0.024, $+$0.271, $+$0.578)   & W2    & - & W1 & 62.24\%    & \checkmark    & 16\\
($+$0.271, $+$0.024, $+$0.578)  & W2    & - & ($+$0.024, $-$0.271, $+$0.578)   & W2    & - & W1 & 62.24\%    & \checkmark    & 16\\
($+$0.271, $+$0.024, $+$0.578)  & W2    & - & ($+$0.024, $+$0.271, $+$0.578)   & W2    & + & W1 & 63.59\%    & $\times$    & 16\\
($+$0.271, $+$0.024, $+$0.578)  & W2    & - & ($-$0.024, $-$0.271, $-$0.578)   & W2    & + & W1 & 67.90\%    & $\times$    & 16\\
($+$0.271, $+$0.024, $+$0.578)  & W2    & - & ($-$0.024, $+$0.271, $-$0.578)   & W2    & - & W1 & 69.14\%    & \checkmark    & 16\\
($+$0.271, $+$0.024, $+$0.578)  & W2    & - & ($+$0.024, $-$0.271, $-$0.578)   & W2    & - & W1 & 69.14\%    & \checkmark    & 16\\
($+$0.271, $+$0.024, $+$0.578)  & W2    & - & ($+$0.024, $+$0.271, $-$0.578)   & W2    & + & W1 & 70.36\%    & $\times$    & 16\\
\hline
($+$0.518, $+$0.014, $+$0.000)  & W1    & + & ($+$0.518, $-$0.014, $+$0.000)   & W1    & - & $\Gamma$ & 2.80\% & $\times$    & 8\\
($+$0.518, $+$0.014, $+$0.000)  & W1    & + & ($-$0.518, $+$0.014, $+$0.000)   & W1    & - & $\Gamma$ & 49.30\% & $\times$    & 8\\
($+$0.518, $+$0.014, $+$0.000)  & W1    & + & ($-$0.518, $-$0.014, $+$0.000)   & W1    & + & $\Gamma$ & 49.38\% & \checkmark    & 8\\
($+$0.518, $+$0.014, $+$0.000)  & W1    & + & ($+$0.014, $+$0.518, $+$0.000)   & W1    & - & $\Gamma$ & 71.28\% & $\times$    & 8\\
($+$0.518, $+$0.014, $+$0.000)  & W1    & + & ($-$0.014, $+$0.518, $+$0.000)   & W1    & + & $\Gamma$ & 73.28\% & \checkmark    & 8\\
($+$0.518, $+$0.014, $+$0.000)  & W1    & + & ($+$0.014, $-$0.518, $+$0.000)   & W1    & + & $\Gamma$ & 73.28\% & \checkmark    & 8\\
($+$0.518, $+$0.014, $+$0.000)  & W1    & + & ($-$0.014, $-$0.518, $+$0.000)   & W1    & - & $\Gamma$ & 75.24\% & $\times$    & 8\\
\end{tabular}
}
\end{table}

\clearpage

\end{bibunit}

\setcounter{figure}{0}
\renewcommand{\thefigure}{S\arabic{figure}}

\clearpage
\onecolumngrid
\begin{figure}[ht]
	\centering
	\includegraphics[width=\linewidth]{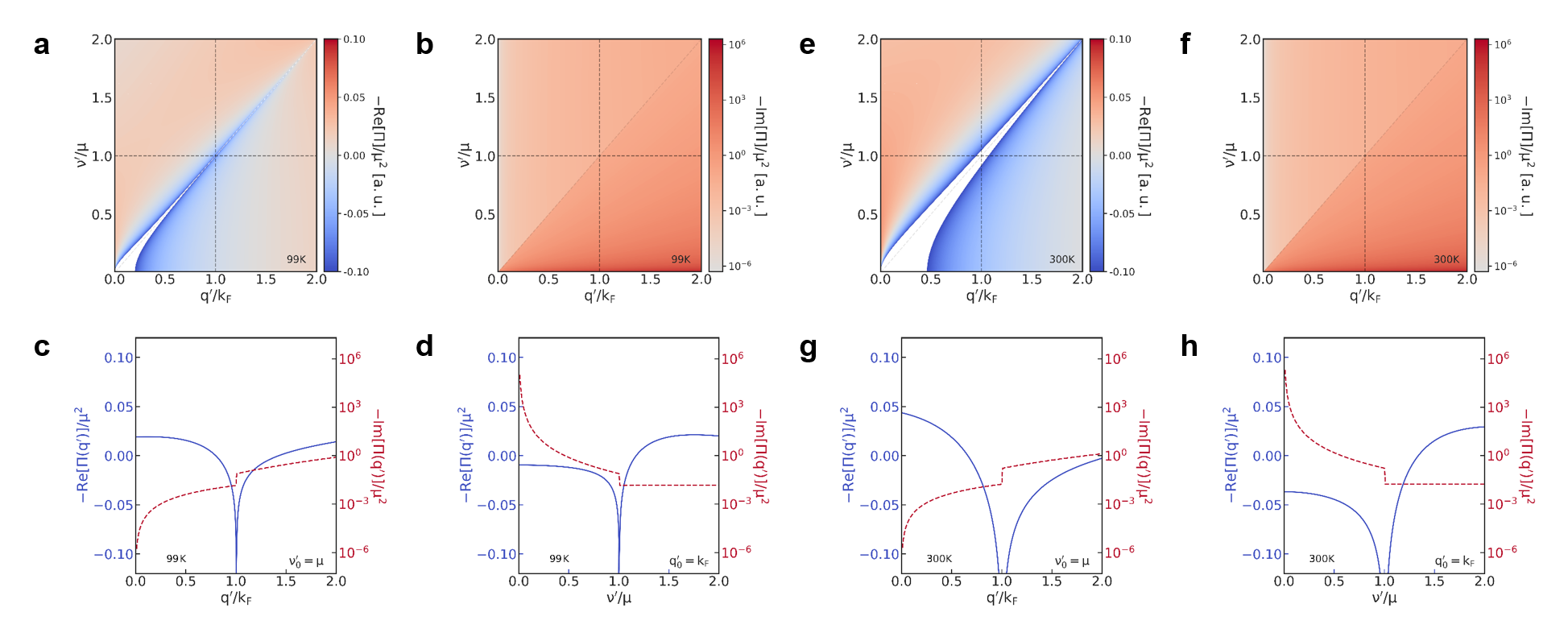}
	\caption{Density plot of the \textbf{a}, real, and the \textbf{b}, imaginary part of the polarization function $-\Pi(\boldsymbol{q}^{\prime}, \nu^{\prime})$ at 99K and with finite doping where $\boldsymbol{q}^{\prime} = \boldsymbol{q} - \boldsymbol{k}_{W1} + \boldsymbol{k}_{W2}$ and $\nu^{\prime} = \nu - \mu_1 + \mu_2$. \textbf{c}. Line profile of negative  $\textrm{Re}[\Pi(\boldsymbol{q}^{\prime}, \nu_0^\prime=\mu)]$ and $\textrm{Im}[\Pi(\boldsymbol{q}^{\prime}, \nu_0^\prime=\mu)]$ at a constant frequency $\nu_0^\prime/\mu=1$. \textbf{d}. Line profile of negative  $\textrm{Re}[\Pi(q_0^\prime = k_F, \nu^{\prime})]$ and $\textrm{Im}[\Pi(q_0^\prime = k_F, \nu^{\prime})]$ at a constant wavevector $q_0^\prime/k_F=1$, where divergence at  $\mathrm{v}_Fq^\prime = \nu^\prime$ is maintained. \textbf{e-h.} Analogous to subfigures \textbf{a-d}, calculated at a temperature of 300K.}
	\label{fig:sup1}
\end{figure}

\clearpage

\begin{figure}[ht]
	\centering
	\includegraphics[width=\linewidth]{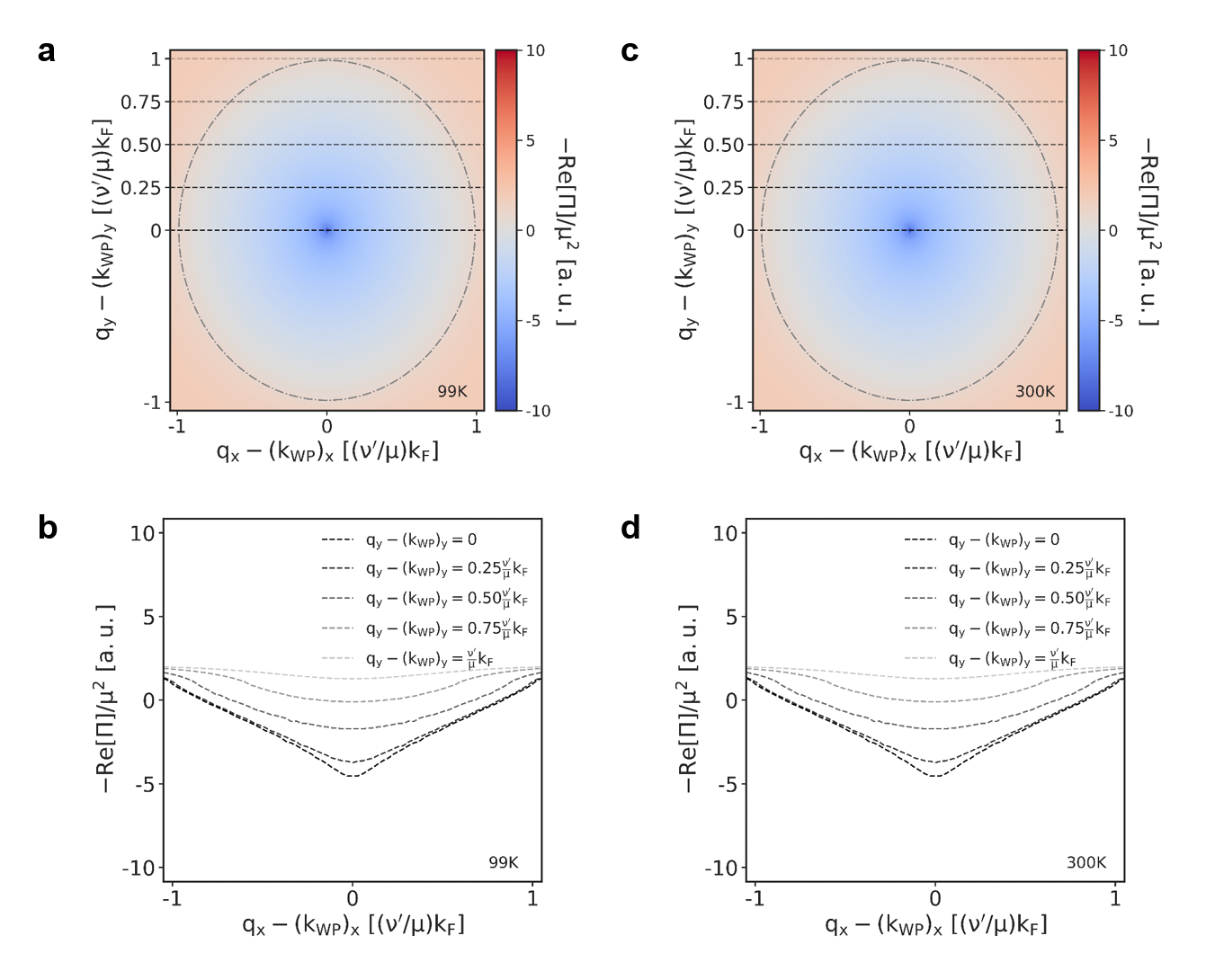}
	\caption{\textbf{a}. Density plot of $-\textrm{Re}[\Pi(\boldsymbol{q_x}-\boldsymbol{k_{\text{WP}_x}}, \boldsymbol{q_y}-\boldsymbol{k_{\text{WP}_y}})]$ integrated over $0 \leq \nu \leq \nu'$ where the divergence is suppressed by additional terms in the denominator representative of other general scattering mechanics. Line cuts at different values of $\boldsymbol{q_y}-\boldsymbol{k_{\text{WP}_y}} = n(\nu'/\mu)k_F$ where $n = 0$, $0.25$, $0.5$, $0.75$ and $1$ are shown in \textbf{b}. \textbf{c-d.} Analogous to subfigures \textbf{a-b}, calculated at a temperature of 300K. The temperature dependence can be seen to be negligible from 1K up to 300K.}
	\label{fig:sup2}
\end{figure}

\clearpage

\begin{figure}[ht]
	\centering
	\includegraphics[width=\linewidth]{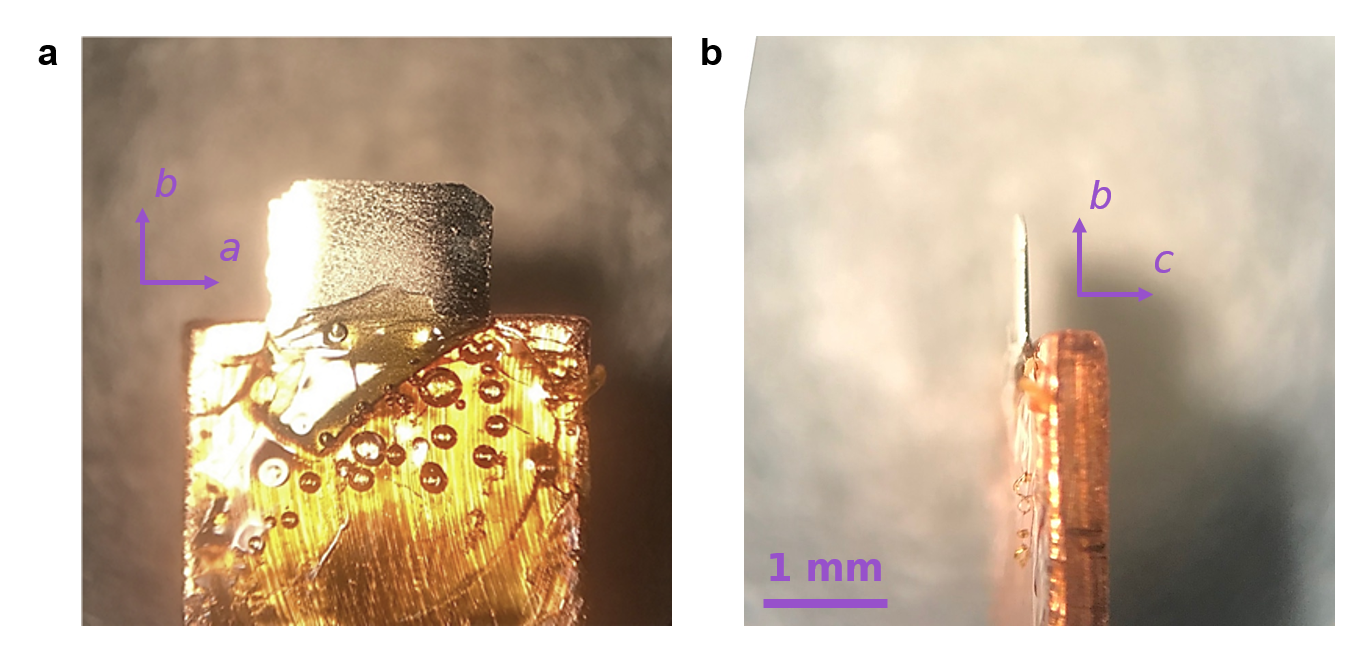}
	\caption{\textbf{a}, Top view and \textbf{b}, side view of the orientated and thinned-down tantalum phosphide (TaP) sample used for the IXS experiments.}
	\label{fig:sup3}
\end{figure}

\clearpage

\begin{figure}[ht]
	\centering
	\includegraphics[width=\linewidth]{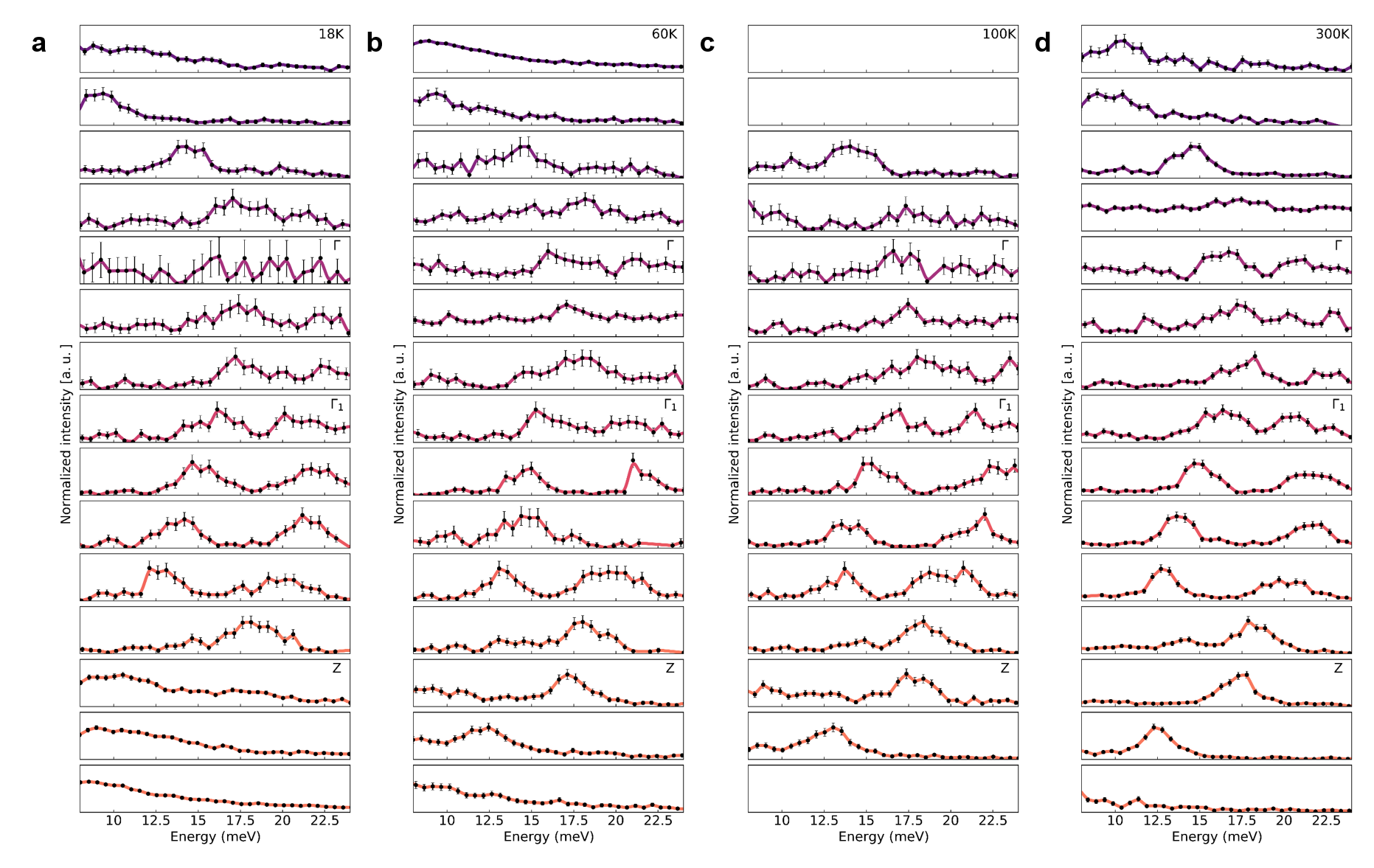}
	\caption{From top to bottom of each subfigure, intensity spectra obtained from IXS measurements performed along high-symmetry loop $\Gamma$-$\Sigma$-$\Sigma_1$-Z-$\Gamma$ at \textbf{a}, 18K, \textbf{b}, 60K, \textbf{c}, 100K and \textbf{d}, 300K. Lines serve as a guide to the eye.}
	\label{fig:sup4}
\end{figure}

\clearpage

\begin{figure}[ht]
	\centering
	\includegraphics[width=\linewidth]{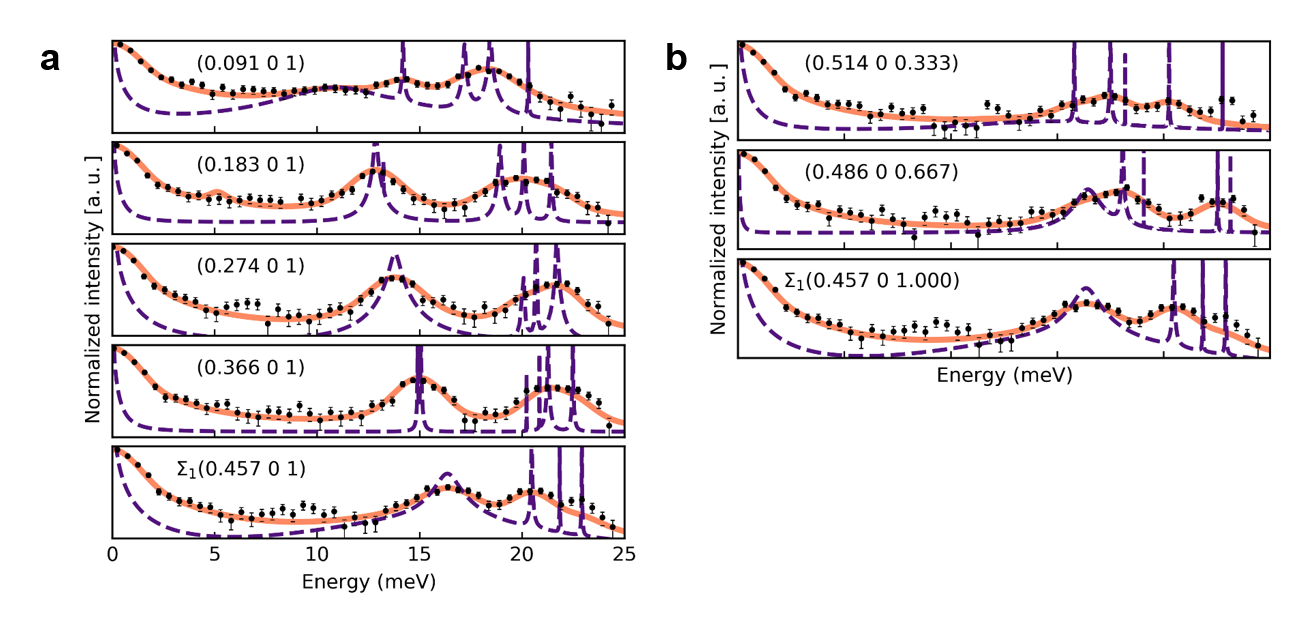}
	\caption{IXS data with statistical error taken at room temperature along high-symmetry line \textbf{a}, $Z-\Sigma_1$ and along \textbf{b}, $\Sigma-\Sigma_1$. The solid orange line denotes a least-squares fit to the intensity spectra from a damped harmonic oscillator function convoluted with the instrumental resolution function. The dashed purple line represents the resolution-deconvoluted phonon modes that contribute to the dispersion relation.}
	\label{fig:sup5}
\end{figure}

\clearpage

\begin{figure}[ht]
	\centering
	\includegraphics[width=\linewidth]{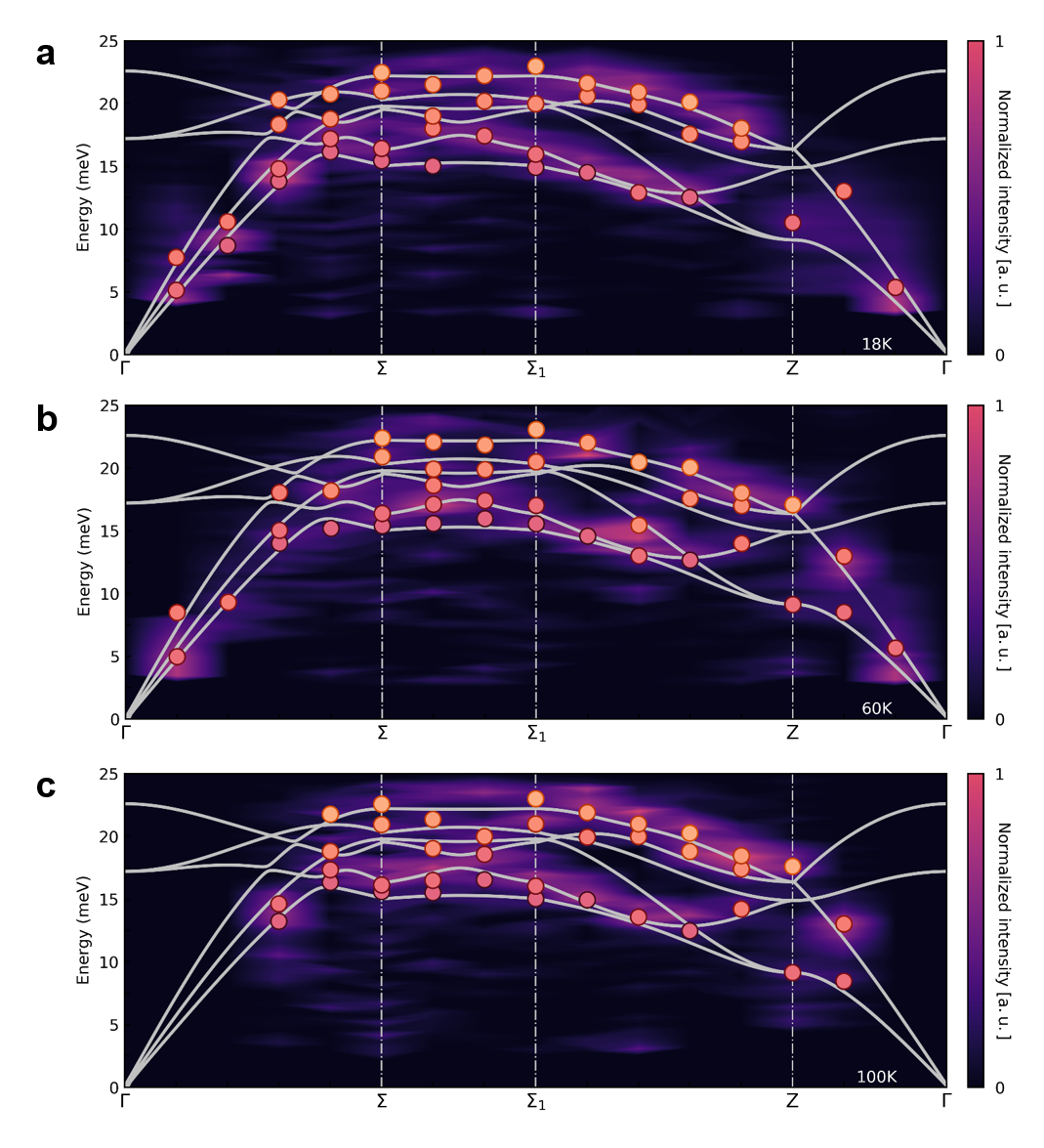}
	\caption{Experimental phonon dispersion of TaP encompassing the three acoustic phonon modes and the three lowest-energy optical modes along high symmetry loop $\Gamma$-$\Sigma$-$\Sigma_1$-Z-$\Gamma$ at \textbf{a}, 18K, \textbf{b}, 60K and \textbf{c}, 100K. The points represent extracted phonon modes from intensity spectra at that $\boldsymbol{q}$-value in momentum space. The grey lines are \textit{ab initio} calculations of the phonon dispersion. The color map shows the intensities of the phonon modes extracted from the intensity spectra following removal of the elastic peak. The first two and the last $\boldsymbol{q}$-values along the high-symmetry loop of the 100K phonon dispersion were not measured.}
	\label{fig:sup6}
\end{figure}

\clearpage

\begin{figure}[ht]
	\centering
	\includegraphics[width=\linewidth]{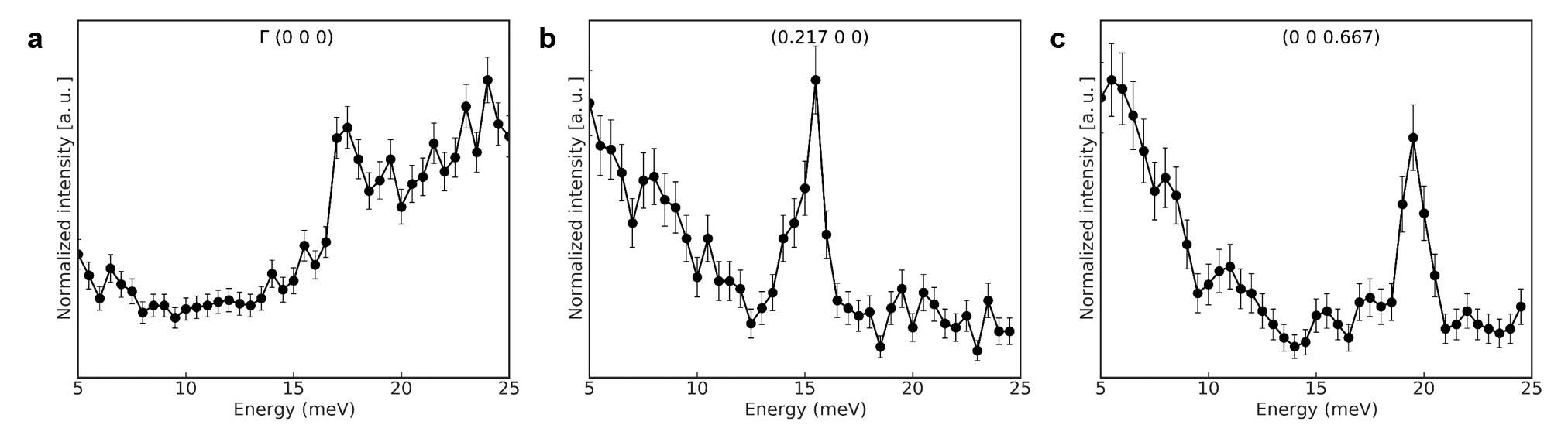}
	\caption{Representative intensity spectra obtained from inelastic neutron scattering (INS) performed with an energy range of 5-25 meV \textbf{a}, at the $\Gamma$ point, \textbf{b}, at a point along high symmetry direction $\Gamma$-$\Sigma$, $(0.217,0,0)$ and \textbf{c}, at a point along high symmetry direction $\Gamma$-Z, $(0,0,0.667)$. These spectra are in good agreement of phonon measurements performed with inelastic x-ray scattering. The solid lines are a guide for the eye.}
	\label{fig:sup7}
\end{figure}

\clearpage

\begin{figure}[ht]
	\centering
	\includegraphics[width=\linewidth]{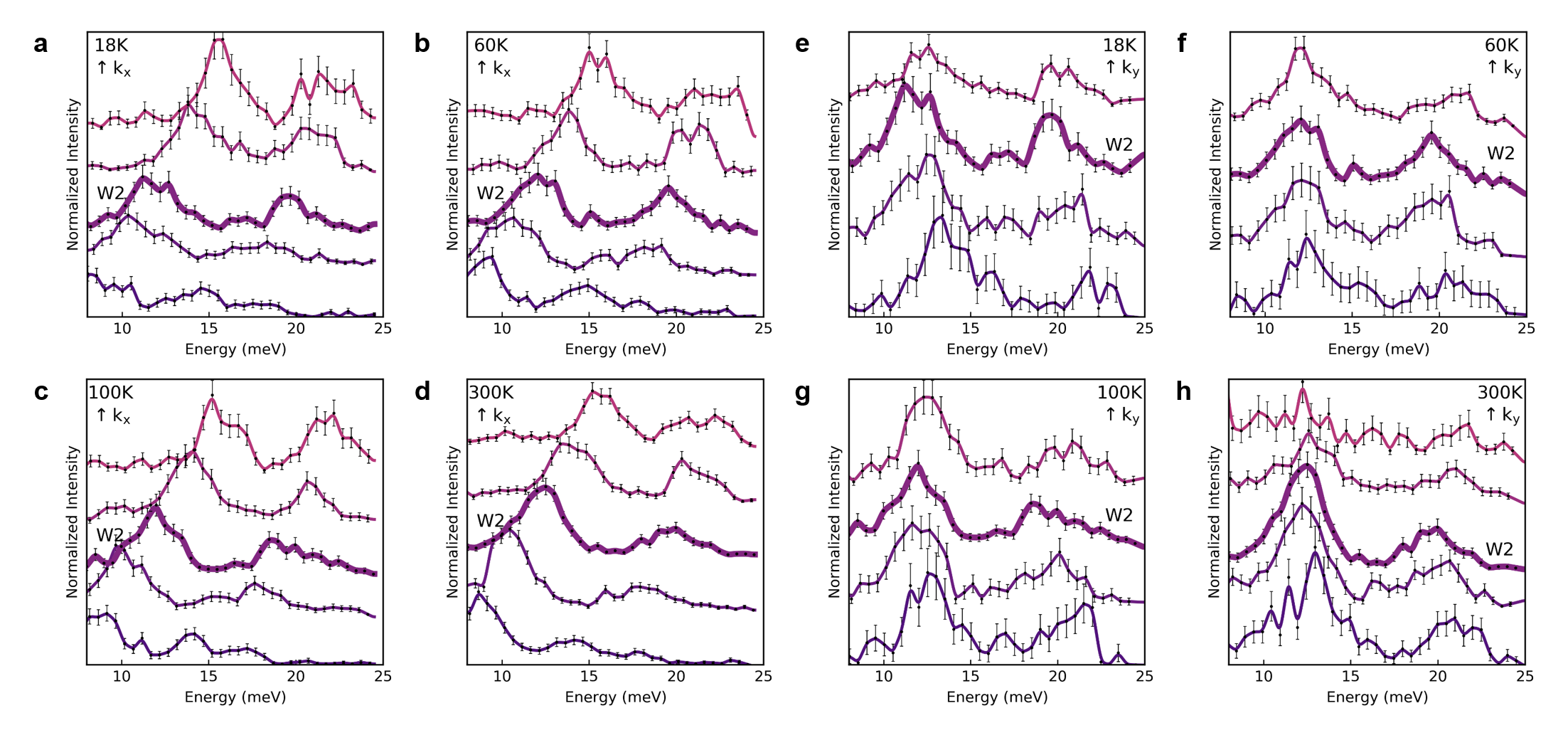}
	\caption{Momentum dependence of the intensity spectra with statistical error near the W2 Weyl node at $(0.271,0.024,0.578)$ (denoted as $\boldsymbol{k}_{W2^{\prime}}$ in main text). \textbf{a-d} display the IXS intensity spectra for $\boldsymbol{q}$-values along $H$ in $(H,0.024,0)$ for increasing $k_x$ direction from bottom to top at 18K, 60K, 100K and 300K, respectively. \textbf{e-h} display similar measurements at the same temperatures for $\boldsymbol{q}$-values along $K$ in $(0.271,K,0)$ for increasing $k_y$ direction from bottom to top. There are only four measurements due to instrumental limitations for larger $\boldsymbol{q}$-values, except for 300K. Lines serve as a guide to the eye. The central thicker line in each of the subfigures represents the IXS measurement performed directly at the W2 node.}
	\label{fig:sup8}
\end{figure}

\clearpage

\begin{figure}[ht]
	\centering
	\includegraphics[width=\linewidth]{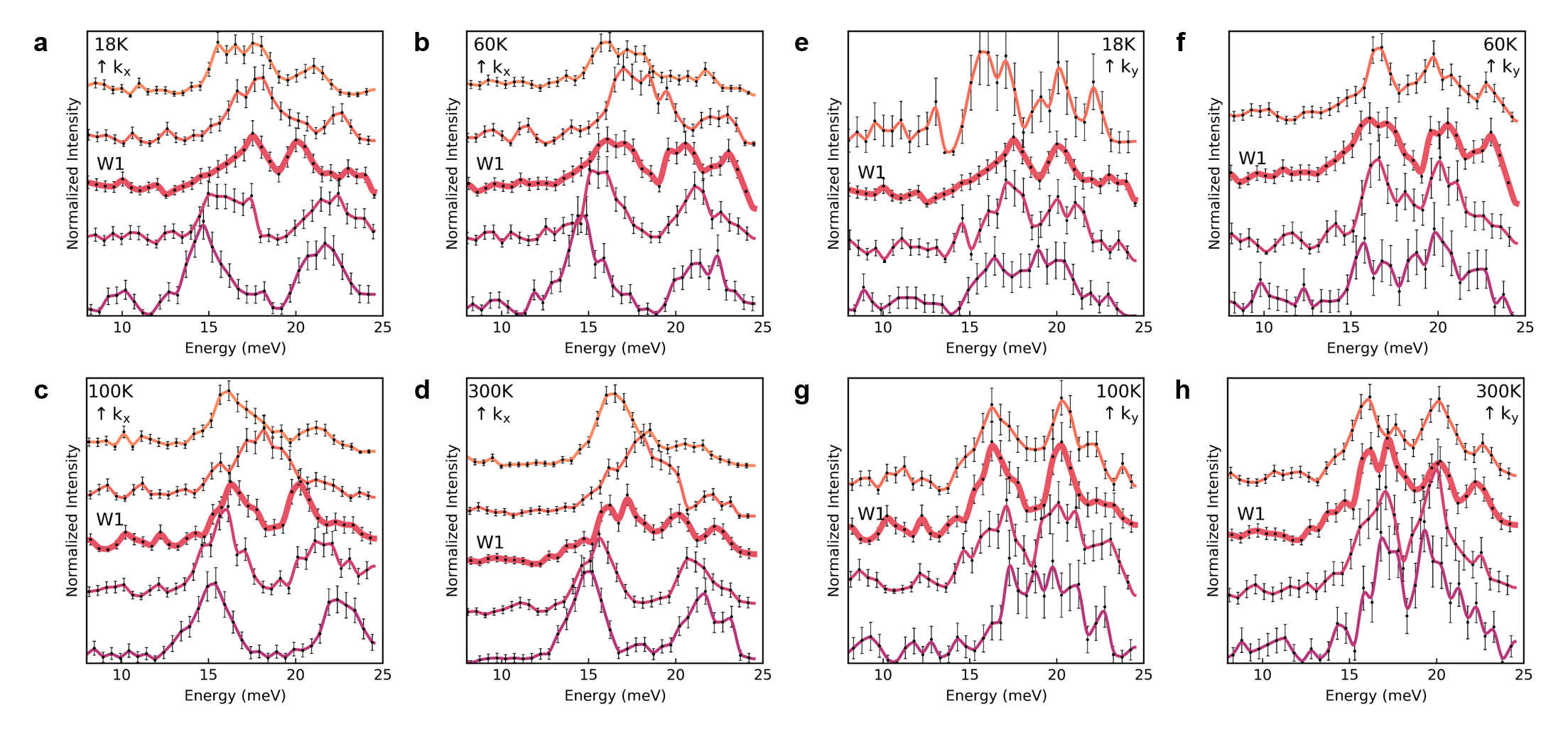}
	\caption{Momentum dependence of the intensity spectra with error bars near the W1 Weyl node at $(0.518,0.014,0)$. \textbf{a-d} display the IXS intensity spectra for $\boldsymbol{q}$-values along $H$ in $(H,0.014,0)$ for increasing $k_x$ direction from bottom to top at 18K, 60K, 100K and 300K, respectively. \textbf{e-h} display similar measurements at the same temperatures for $\boldsymbol{q}$-values along $K$ in $(0.518,K,0)$ for increasing $k_y$ direction from bottom to top. There are only four measurements due to instrumental limitations for larger $\boldsymbol{q}$-values. Lines serve as a guide to the eye. The central thicker line in each of the subfigures represents the IXS measurement performed directly at the W1 node.}
	\label{fig:sup9}
\end{figure}

\clearpage

\begin{figure}[ht]
	\centering
	\includegraphics[width=\linewidth]{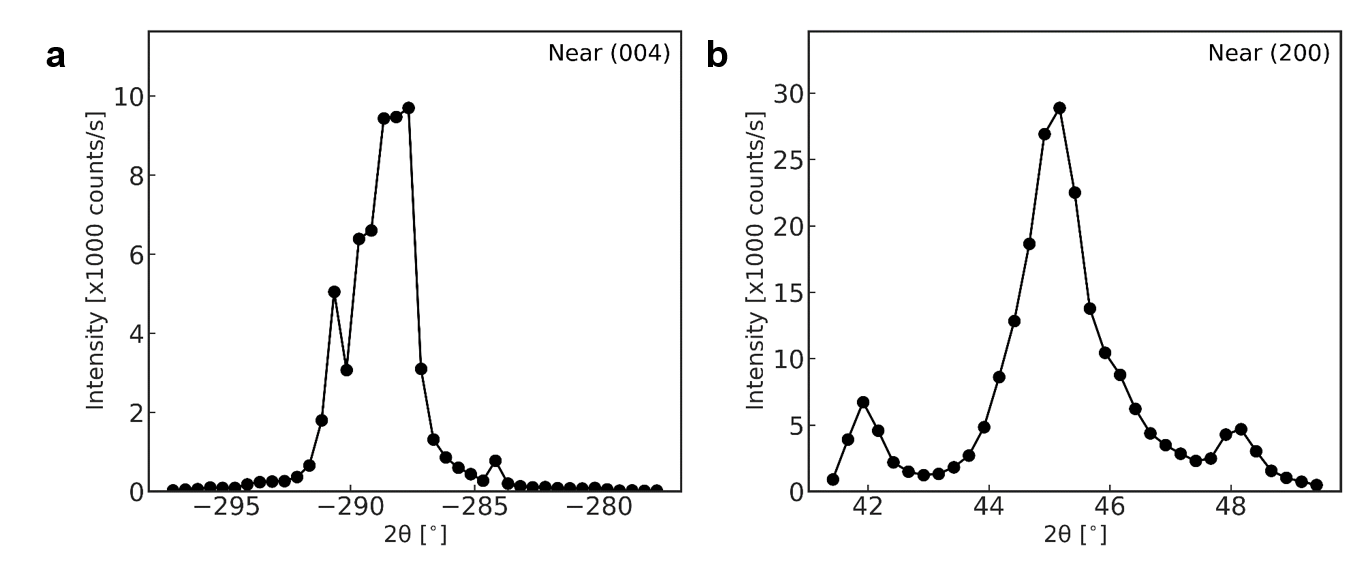}
	\caption{2$\theta$-scans obtained from neutron diffraction scans of single-crystal TaP near Bragg peaks \textbf{a,} (004) and \textbf{b,} (200) at room temperature. Error bars representing one standard deviation are too small to be discernible.}
	\label{fig:sup10}
\end{figure}

\clearpage

\begin{figure}[ht]
	\centering
	\includegraphics[width=\linewidth]{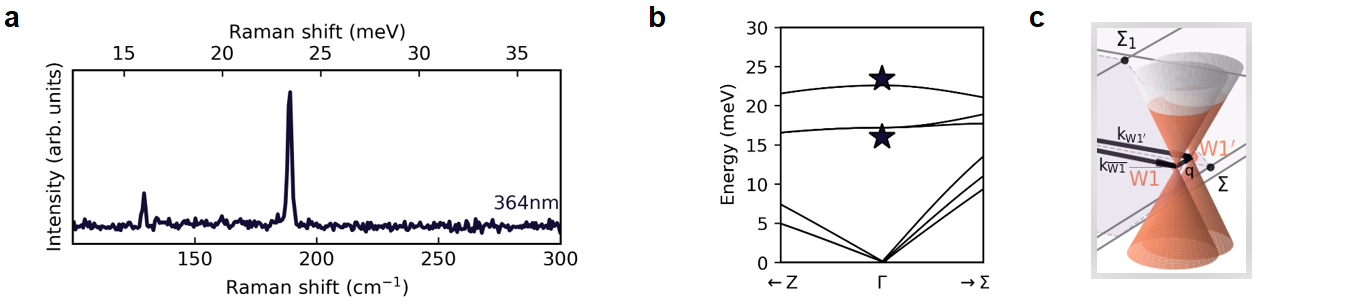}
	\caption{ \textbf{a}. Raman peak shifts of TaP collected at room temperature at laser excitation 364nm. \textbf{b}. Comparison between Raman-measured phonon modes at the $\Gamma$ point and \textit{ab initio} calculations, which is the same dispersion seen in Figure \hyperref[fig:1]{1f}.  \textbf{c}. Close-up of the two W1 nodes near the zone boundary that connect as a small $\boldsymbol{q}\approx \Gamma$ vector.}
	\label{fig:sup11}
\end{figure}

\end{document}